\documentclass[twocolumn,showpacs,eqsecnum,aps]{revtex4}
\usepackage{graphicx}
\usepackage{dcolumn}
\usepackage{amsmath}
\makeatletter
\def\btt#1{\texttt{\@backslashchar#1}}%
\DeclareRobustCommand\bblash{\btt{\@backslashchar}}%
\makeatother
\begin{document}
\title{\ \ \ \\
       \vspace{-1.7cm}
       \hfill SAGA-HE-170-01, TMU-NT-01-01 \\
                           \ \\
                           \ \\
        Determination of nuclear parton distributions}
\author{M. Hirai, S. Kumano}
\homepage{http://www-hs.phys.saga-u.ac.jp}
\email{hirai@hsg.phys.saga-u.ac.jp, kumanos@cc.saga-u.ac.jp}
\affiliation{Department of Physics, Saga University \\
         Saga, 840-8502, Japan}
\author{M. Miyama}
\email{miyama@comp.metro-u.ac.jp}
\affiliation{Department of Physics, Tokyo Metropolitan University \\
         Tokyo, 192-0397, Japan}         
\date{March 19, 2001}
\begin{abstract}
\vspace{0.2cm}
Parametrization of nuclear parton distributions is investigated
in the leading order of $\alpha_s$. 
The parton distributions are provided at $Q^2$=1 GeV$^2$ with
a number of parameters, which are determined by a $\chi^2$ analysis
of the data on nuclear structure functions.
Quadratic or cubic functional form is assumed for the initial
distributions. Although valence quark distributions in the medium $x$
region are relatively well determined, the small $x$ distributions depend
slightly on the assumed functional form. It is difficult to determine
the antiquark distributions at medium $x$ and gluon distributions.
From the analysis, we propose parton distributions at $Q^2$=1 GeV$^2$ 
for nuclei from deuteron to heavy ones with the mass number $A\sim 208$.
They are provided either analytical expressions or computer subroutines
for practical usage. Our studies should be important for understanding
the physics mechanism of the nuclear modification and also for
applications to heavy-ion reactions. This kind of nuclear parametrization
should also affect existing parametrization studies in the nucleon
because ``nuclear" data are partially used for obtaining the optimum
distributions in the ``nucleon".
\end{abstract}
\maketitle
\tableofcontents

\section{Introduction}\label{intro}
\setcounter{equation}{0}

Unpolarized parton distributions in the nucleon are now well determined
in the region from very small $x$ to large $x$ by using various experimental
data. There are abundant data on electron and muon deep inelastic scattering.
In addition, there are available data from neutrino reactions,
Drell-Yan processes, $W$ productions, direct photon productions, and others.
The optimum parton distributions are determined so as to fit these
experimental data. Initial distributions are assumed at 
a fixed $Q^2$ with parameters, which are determined by a $\chi^2$ analysis. 
Now, there are three major groups, 
CTEQ (Coordinated Theoretical/Experimental Project on QCD Phenomenology
and Tests of the Standard Model) \cite{cteq5},
GRV (Gl\"uck, Reya, and Vogt) \cite{grv98}, and
MRS (Martin, Roberts, and Stirling) \cite{mrs99},
which have been investigating the unpolarized parametrization. 

Because the distributions themselves are associated with nonperturbative
Quantum Chromodynamics (QCD), they cannot be described precisely
by theoretical methods at this stage. Therefore, the determination
enables us to understand internal structure of the nucleon, hence
to test the nonperturbative hadron models.
In addition, the parton distributions are always necessary for
calculating cross sections of high-energy reactions involving a nucleon.
Furthermore, there is other importance in determining future
physics direction. If the distributions are precisely known,
any experimental deviation from theoretical predictions should indicate
new physics beyond the standard model, for example a signature of
subquark system.

The situation is worse in polarized distributions.
Longitudinally polarized distributions have been investigated
in the last decade. From the analyses of many polarized electron
and muon deep inelastic experimental data, several parametrizations
have been proposed.
Neutrino, Drell-Yan, and other data are not available
unlike the unpolarized studies, so that antiquark and
gluon distributions cannot be determined accurately 
at this stage \cite{aac}. More precise determination should 
be done by the polarized Relativistic Heavy Ion Collider (RHIC)
experiments in the near future.

It is well known that nuclear parton distributions are modified from
the corresponding ones in the nucleon according to the measurements
of nuclear $F_2$ structure functions \cite{f2sum}. 
Because $F_2$ is expected to be dominated by sea- and valence-quark
distributions at small and large $x$, respectively, the modification
of each distribution should be determined by the
$F_2$ measurements in the small or large $x$ region. However, it is 
not straightforward to find the detailed $x$ dependence of these
distributions. Furthermore, it is also not obvious how nuclear
gluon distributions are constrained by the $F_2$ data.
There are trials to obtain the distributions from the data
in a model dependent way \cite{saga} and in a model independent
way by Eskola, Kolhinen, and Ruuskanen \cite{helsinki}; however,
they are not $\chi^2$ analyses. Therefore, this paper is intended
to pioneer the $\chi^2$ analysis study for obtaining optimum
nuclear parton distributions \cite{mumbai}.

There are following motivations for investigating the nuclear
parametrization. The first purpose is to test various nuclear models
for describing the nuclear structure functions. From the detailed
comparison, the most appropriate nuclear model could be determined
in the high-energy region. Furthermore, we may be able to find
an explicit subnucleon signature in nuclear physics.
The second purpose is to calculate cross sections of high-energy
nuclear reactions accurately.
For example, precise initial conditions must be known in heavy-ion
reactions for finding a signature of quark-gluon plasma.
In many calculations, nuclear parton distributions are simply assumed
to be the same as the corresponding ones in the nucleon.
The third purpose is related to the aforementioned nucleon
parametrization. In obtaining parton distributions in the ``nucleon",
some ``nuclear" data are actually used without considering
the nuclear modification.
For example, neutrino $F_3$ data are essential for determining
the valence-quark distributions in the nucleon. However, most 
data are taken for the iron target! In future, accurate neutrino-proton
scattering data could be taken if a neutrino factory is realized
\cite{sknu}. It is inevitable to utilize the nuclear data at this stage. 
Therefore, our nuclear studies should be useful for improving
the present parametrizations in the nucleon.

This paper consists of the following.
First, dependence of the mass number $A$ and scaling
variable $x$ is discussed in \ref{ax-depend}.
Then, our analysis method is explained in \ref{method}, and results
are presented in \ref{results}. Obtained distributions
are provided in Sec. \ref{usage} for practical usage.
Finally, our nuclear parametrization studies are summarized
in section \ref{sum}.

\section{Functional form of nuclear parton distributions}
\label{ax-depend}
\setcounter{equation}{0}

Because there is no prior $\chi^2$ analysis on the nuclear parametrization,
we should inevitably take the $A$ and $x$ dependence as simple as possible 
for the first step trial. However, the functional form of $x$ should be
taken independently from any theoretical nuclear models as a fair analysis.
Our standpoint is to test the models without relying on them.
We discuss an appropriate functional form of the distributions
in the following subsections.

\subsection{$A$ dependence}
\label{a-depend}

There is some consensus on physics mechanism of nuclear modification
in each $x$ region. Because different mechanisms contribute to the 
modification depending on the $x$ region, the $A$ dependence could
vary in different $x$ regions. However, in order to simplify the analysis,
we introduce a rather bold assumption:
The $A$ dependence is assumed to be proportional to $1/A^{1/3}$.
Physics behind this idea is discussed by Day and Sick in Ref. \cite{ds}.
In any nuclear reaction, the cross section is expressed in terms
of nuclear volume and surface contributions:
\begin{equation}
\sigma_A = A \, \sigma_V + A^{2/3} \sigma_S.
\end{equation}
Therefore, the cross section per nucleon is given as
\begin{equation}
\frac{\sigma_A}{A} = \sigma_V + \frac{1}{A^{1/3}} \sigma_S.
\end{equation}
If $\sigma_V$ and $\sigma_S$ depend weakly on $A$,
the $1/A^{1/3}$ dependence makes sense as the leading approximation.
In fact, according to the measured $F_2^A/F_2^D$ data, this $A$ dependence
is justified \cite{ds}. 

\begin{figure}[h]
\includegraphics[width=0.46\textwidth]{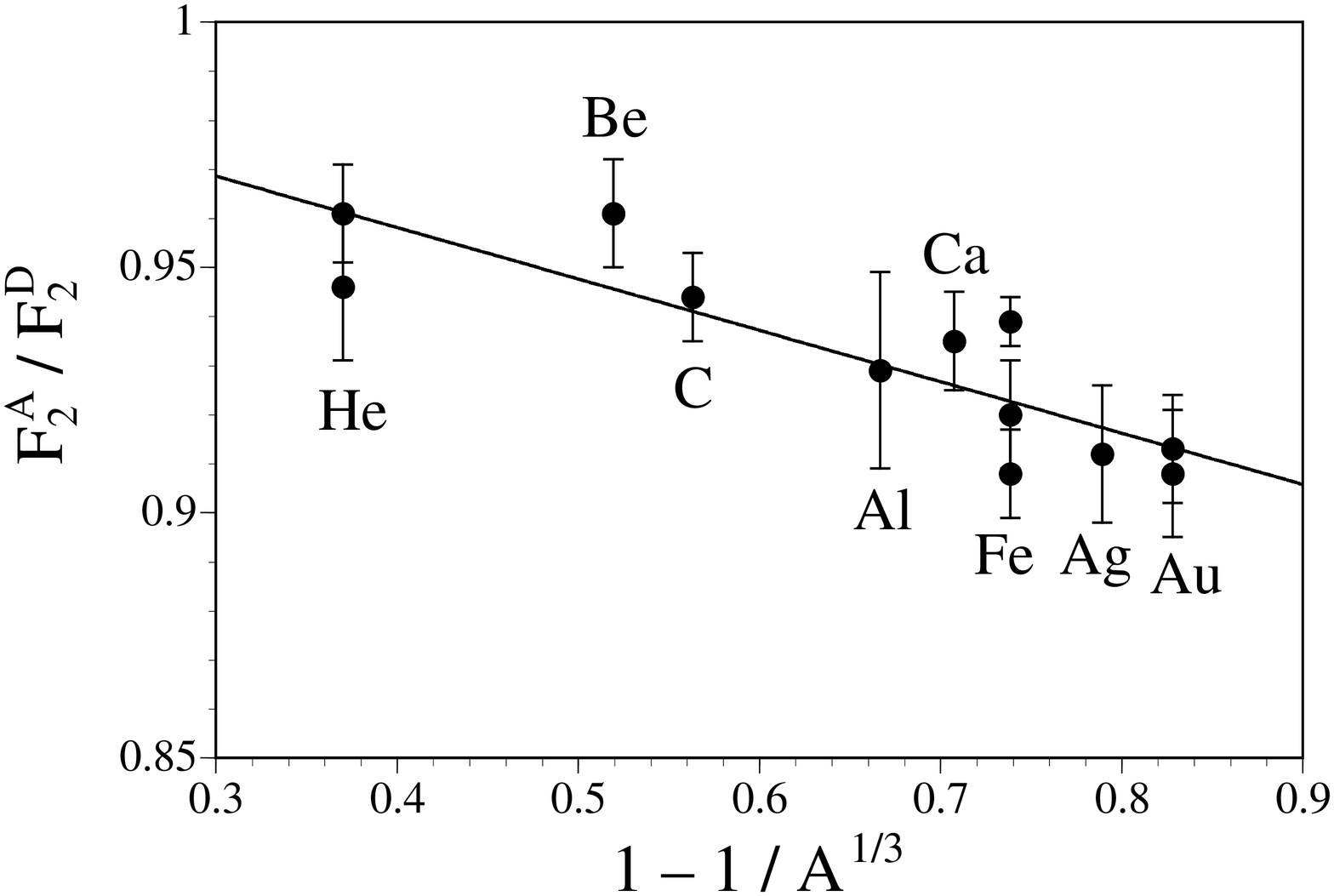}
\vspace{-0.2cm}
\caption{$A$ dependence of measured $F_2^A/F_2^D$ ratios
         at $x=0.5$.}
\label{fig:adep}
\end{figure}
In order to illustrate the justification, we show actual
$F_2^A/F_2^D$ data as a function of $1-1/A^{1/3}$ at $x=0.5$
together with a straight line in Fig. \ref{fig:adep}. 
It is obvious from the figure that measured ratios are well 
reproduced by the line with the $1-1/A^{1/3}$ dependence.
We also looked at other regions, for example, the small $x$ region
and found that the data could be described by this $A$ dependence.
In this way, we decided to employ this simple assumption.
Of course, the $1/A^{1/3}$ dependence may be an oversimplification.
Detailed studies of the $A$ dependence are left as a future
research topic.

We should mention, however, that the exact $1-1/A^{1/3}$ dependence
is not completely consistent with the conditions of nuclear
charge, baryon number, and momentum. We need to assign fine-tuning
parameters which adjust the nuclear dependence. 
The details will be explained in the next subsection
below Eq. (\ref{eqn:wi-withh}).

\subsection{$x$ dependence}
\label{x-depend}

Available nuclear data on the structure function $F_2^A$ are 
taken in fixed-target experiments at this stage,
and they are shown as a function of $Q^2$ and $x \equiv Q^2/(2m\nu)$,
where $\nu$ is the energy transfer, $m$ is the nucleon mass, and 
$Q^2$ is given by $Q^2 \equiv - q^2$ with the virtual photon momentum $q$.
The initial nuclear parton distributions are provided
at a fixed $Q^2$ ($\equiv Q_0^2$), and they are taken as
\begin{equation}
f_i^A (x, Q_0^2) = w_i(x,A,Z) \, f_i (x, Q_0^2),
\label{eqn:apart}
\end{equation}
where $f_i^A (x, Q_0^2)$ is the parton distribution with type $i$
in the nucleus $A$ and $f_i (x, Q_0^2)$ is the corresponding parton
distribution in the nucleon. We call $w_i(x,A,Z)$ a weight function,
which takes into account the nuclear modification. 

We lose a piece of information by using Eq. (\ref{eqn:apart}).
A scaling variable for the lepton-nucleus scattering could be given by
$x_A \equiv Q^2/(2p_A\cdot q)$, where $p_A$ is the momentum of the target
nucleus. We could define another variable for lepton-nucleon scattering
as $x_N \equiv Q^2/(2p\cdot q)$ with the nucleon momentum $p$ 
within the nucleus. 
The variable $x_A$ indicates the momentum fraction of a struck-quark
in the nucleus and it is kinematically restricted as $0 < x_A < 1$.
Because of these definitions, we have the relation $x=x_A M_A/m$,
where $M_A$ is the nuclear mass, in a fixed target reaction.
In this way, we find that the range of nuclear parton distributions
is given by $0<x<A$. Therefore, the extremely large $x$ region
($x>1$) cannot be described in our present approach
with Eq. (\ref{eqn:apart}) because the distributions
in the nucleon vanish: $f_i (x>1, Q_0^2)=0$. 

There are following reasons for using Eq. (\ref{eqn:apart})
irrespective of this issue.
First, there are not so many experimental data in this $x$ region.
Second, even if finite parton distributions are assumed at the
initial point $Q_0^2$, there is no reliable theoretical tool to evolve
them to larger $Q^2$ points where the data exist. Therefore, such
large $x$ distributions cannot be accommodated in our $\chi^2$ analysis.
There is a nuclear $Q^2$ evolution code in Ref. \cite{bf1};
however, the original evolution equations are written
in the range $0<x<1$ \cite{mq}.
Third, because the nuclear modification is generally in the 10$-$30\%
range for medium and large size nuclei, 
it is much easier to investigate the modification from
the distributions in the nucleon rather than the absolute nuclear
distributions themselves.
Forth, the structure functions $F_2^A$ or parton distributions are tiny
at $x>1$, so that they do not affect the calculated cross sections
significantly unless an extreme kinematical condition is chosen.

Now, we proceed to the actual $x$ dependence.

\noindent
{\bf (1) Nuclear modification $\bf \propto$ $\bf 1-1/A^{1/3}$}

As concluded in the previous subsection, the nuclear modification
part of $w_i (x,A,Z)$ is assumed to be proportional to $1-1/A^{1/3}$:
\begin{equation}
w_i(x,A,Z)=1+\left( 1 - \frac{1}{A^{1/3}} \right) 
            (\text{function of $x$, $A$, and $Z$}). 
\label{eqn:wia}
\end{equation}

\noindent
{\bf (2) Introduction of $\bf 1/(1-x)^\beta$ factor}
      
The nuclear parton distributions have finite values
even at $x=1$ in principle, so that the weight function should behave as
\begin{equation}
w_i (x\rightarrow 1,A,Z)  = 
    \frac{f_i^A(x \rightarrow 1)}{f_i (x \rightarrow 1)}
   \rightarrow \infty.
\end{equation}
This equation should hold whatever the modification mechanism is.
In order to reproduce this feature, the factor $1/(1-x)^{\beta_i}$
is introduced in the $x$ dependent function part of Eq. (\ref{eqn:wia})
with a parameter $\beta_i (>0)$:
\begin{equation}
(\text{function of $x$, $A$, and $Z$}) \propto \frac{1}{(1-x)^{\beta_i}}.
\end{equation}

\noindent
{\bf (3) Saturation of shadowing or antishadowing}

We assume saturation of the function $w_i(x,A,Z)$ at $x \rightarrow 0$.
It is considered to be a reasonable assumption unless
there is a peculiar mechanism to produce singular behavior
at small $x$. As far as the experimental $F_2^A/F_2^D$ data suggest,
the shadowing at small $x$ tends to saturate as $x\rightarrow 0$.
If this saturation is assumed, the weight function $w_i$ becomes
\begin{equation}
w_i(x \rightarrow 0,A,Z)=1+\left( 1 - \frac{1}{A^{1/3}} \right) a_i(A,Z),
\end{equation}
where $a_i$ is the parameter which controls shadowing or
antishadowing magnitude.

From these discussions, an appropriate functional form becomes
\begin{equation}
w_i(x,A,Z)=1+\left( 1 - \frac{1}{A^{1/3}} \right) 
          \frac{a_i(A,Z)+H_i(x)}{(1-x)^{\beta_i}},
\label{eqn:wi-withh}
\end{equation}
where the additional $x$ dependent part is written as $H_i(x)$,
and it determines minute $x$ dependent shape.
The reason why we let the parameters $a_i$ depend on $A$ and $Z$
is the following. As explained in the next section,
there are three obvious constraints, charge, baryon number, and
total momentum, on the nuclear distributions. 
If these conditions are satisfied for a nucleus, they are also
fulfilled for other nuclei with the same $Z/A$ ratio.
For example, the conditions could be satisfied for
all the isoscalar nuclei. 
However, we also analyze nuclei with different $Z/A$ ratios.
Then, even if the three conditions are satisfied in a certain
nucleus, they are not strictly fulfilled for other nuclei with
different $Z/A$ factors.
In order to avoid this kind of failure, nuclear dependence
is assumed for the parameters $a_i$ in the valence-quark and 
gluon functions. They are determined by the $\chi^2$ analysis
so as to satisfy the three conditions for any nucleus.

Because this is the first $\chi^2$ trial, we would like to simplify
the functional form of $H_i(x)$ as much as possible.
A simple idea is to expand it in a polynomial form
$H_i(x)=b_i x +c_i x^2 + \cdot\cdot\cdot$.
An advantage of this functional form is that the polynomials
of $x$ and $1-x$ are very easy to be handled
in the Mellin-transformation method of the $Q^2$ evolution.
Because a direct $x$-space solution \cite{bf1} is used in our $Q^2$
evolution, it does not matter in our present analysis.
However, it is important for public usage if we consider
the fact that many researchers use the Mellin transformation
as their $Q^2$ evolution method. 

It is obvious that $H_i (x)=b_i x$ cannot explain 
the complicated $F_2^A/F_2^D$ shape in the medium $x$ region.
Therefore, the simplest yet realistic choice is
to take $H_i(x)=b_i x+c_i x^2$. This seems to be acceptable
in the sense that the medium $x$ depletion of $F_2^A/F_2^D$
can be explained together with the $F_2^A$ shadowing at small $x$.
However, if the depletion is described by the valence-quark behavior
as our common sense suggests, the valence-quark distributions show
anti-shadowing at small $x$ due to the baryon-number conservation.
As explained in Refs.\cite{F3A,sknu}, it is not obvious at this stage
whether the valence distributions indicate shadowing or antishadowing
without accurate neutrino-deuteron scattering data and also small $x$
nuclear data. Therefore, this functional form should be considered
as one of possible options:
\begin{align}
w_i(x,A,Z) & =1+\left( 1 - \frac{1}{A^{1/3}} \right) 
          \frac{a_i(A,Z) +b_i x+c_i x^2}{(1-x)^{\beta_i}}
\nonumber \\
    & \text{``quadratic type"}.
\label{eqn:quad-wi}
\end{align}
We call this weight function the quadratic type in the following
discussions. 

If the next polynomial term $d_i x^3$ is added in addition,
the function becomes more flexible in fitting the data:
\begin{align}
w_i(x,A,Z) & = 1+\left( 1 - \frac{1}{A^{1/3}} \right) 
          \frac{a_i(A,Z) +b_i x+c_i x^2 +d_i x^3}{(1-x)^{\beta_i}}
\nonumber \\
      & \text{``cubic type"}.
\end{align}
We call this weight function the cubic type.
An advantage of the additional term is, for example, that 
the weight function becomes flexible enough to accommodate
both possibilities, shadowing and anti-shadowing,
in the valence-quark distributions.
A disadvantage is that the number of total parameters becomes larger,
so that the total computing time becomes longer.

These quadratic and cubic functional types are used in our $\chi^2$
analyses. Finding the optimum point for the parametrization set,
we expect to explain the major features of the measured $F_2^A/F_2^D$
ratios.

\section{Analysis method}
\label{method}
\setcounter{equation}{0}

In addition to the initial functional form,
there are other important factors for performing the $\chi^2$ analysis. 
In this section, the details are discussed on used experimental data
and our $\chi^2$ analysis method.

\subsection{Experimental data}
\label{data}

There are many available experimental data which could indicate 
nuclear parton distributions. However, we restrict the data to those
taken by the deep inelastic electron and muon scattering.
Neutrino, Drell-Yan, and other hadron-collider data are not used
in our present analysis with the following reasons.
At first, we would like to investigate how the distributions
are determined solely by the electron and muon scattering data.
Next, as it is mentioned in Sec. \ref{intro}, nuclear modification of
$F_3$ is not measured in the neutrino scattering. 
In hadron-hadron reactions, there are also issues of $K$-factors 
and final state interactions. In a future version of our analysis,
we will consider to include other data. 

\begin{table}[t!]
\caption{Nuclear species, references, and data numbers are listed
         for the used experimental data with $Q^2 \ge 1$ GeV$^2$.}
\label{tab:exp}
\begin{ruledtabular}
\begin{tabular*}{\hsize}
{c@{\extracolsep{0ptplus1fil}}l@{\extracolsep{0ptplus1fil}}c
@{\extracolsep{0ptplus1fil}}c}
nucleus & experiment & reference & \# of data \\
\colrule
He
        & SLAC-E139 & \cite{slac94}    &     18          \\
        & NMC-95    & \cite{nmc95}     &     17          \\
Li      & NMC-95    & \cite{nmc95}     &     17          \\
Be      & SLAC-E139 & \cite{slac94}    &     17          \\
C
        & EMC-88    & \cite{emc88}     &    \ 9          \\
        & EMC-90    & \cite{emc90}     &    \ 5          \\
        & SLAC-E139 & \cite{slac94}    &    \ 7          \\
        & NMC-95    & \cite{nmc95}     &     17          \\
        & FNAL-E665-95   & \cite{e665-95} & \ 5          \\
N       & BCDMS-85  & \cite{bcdms85}   &    \ 9          \\
Al
        & SLAC-E49  & \cite{slac83B}   &     18          \\
        & SLAC-E139 & \cite{slac94}    &     17          \\
Ca
        & EMC-90    & \cite{emc90}     &    \ 5          \\
        & NMC-95    & \cite{nmc95}     &     16          \\
        & SLAC-E139 & \cite{slac94}    &    \ 7          \\
        & FNAL-E665-95   & \cite{e665-95} & \ 5          \\
Fe
        & SLAC-E87  & \cite{slac83}    &     14          \\
        & SLAC-E140 & \cite{slac88}    &     10          \\
        & SLAC-E139 & \cite{slac94}    &     23          \\
        & BCDMS-87  & \cite{bcdms87}   &     10          \\
Cu      & EMC-93    & \cite{emc93}     &     19          \\
Ag      & SLAC-E139 & \cite{slac94}    &    \ 7          \\
Sn      & EMC-88    & \cite{emc88}     &    \ 8          \\
Xe      & FNAL-E665-92   & \cite{e665-92} & \ 5          \\
Au
        & SLAC-E140 & \cite{slac88}    &    \ 1          \\
        & SLAC-E139 & \cite{slac94}    &     18          \\
Pb      & FNAL-E665-95   & \cite{e665-95} & \ 5          \\
\colrule
total   &           &                  &   309 \ \       \\
\end{tabular*}
\end{ruledtabular}
\end{table}

At this stage, the available nuclear data are mainly on
the $F_2$ structure functions in the electron and muon scattering.
Experimental results are shown by the ratio
\begin{equation}
R_{F_2} ^A (x,Q^2) \equiv \frac{F_2^A (x,Q^2)}{F_2^D (x,Q^2)}.
\end{equation}
Information about the used experimental data is given in
Table \ref{tab:exp}, where nuclear species, references, and data
numbers are listed.
The experimental data are taken from the publications by
the European Muon Collaboration (EMC) \cite{emc88, emc90, emc93}
at the European Organization for Nuclear Research (CERN),
the E49, E87, E139, and E140 Collaborations
\cite{slac83,slac83B,slac88,slac94}
at the Stanford Linear Accelerator Center (SLAC),
the Bologna-CERN-Dubna-Munich-Saclay (BCDMS) Collaboration
\cite{bcdms85, bcdms87} at CERN,
the New Muon Collaboration (NMC) \cite{nmc95} at CERN,
the E665 Collaboration \cite{e665-92,e665-95}
at the Fermi National Accelerator Laboratory (FNAL).
The used data are for the nuclei:
helium (He), lithium (Li), beryllium (Be), carbon (C), nitrogen (N),
aluminum (Al),  calcium (Ca), iron (Fe), copper (Cu), silver (Ag), tin (Sn), 
xenon (Xe), gold (Au), and lead (Pb).
As explained in the next subsection, $Q_0^2$=1 GeV$^2$ is taken
as the point where the parton distributions are defined.
Because the published data include smaller $Q^2$
points, it is necessary to choose the ones with large enough $Q^2$
which could be considered in the perturbative QCD region.
The only data with $Q^2 \ge 1$ GeV$^2$ are taken
into account in the $\chi^2$ analysis. The total number of the data
is 309.

\begin{figure}[t!]
\includegraphics[width=0.46\textwidth]{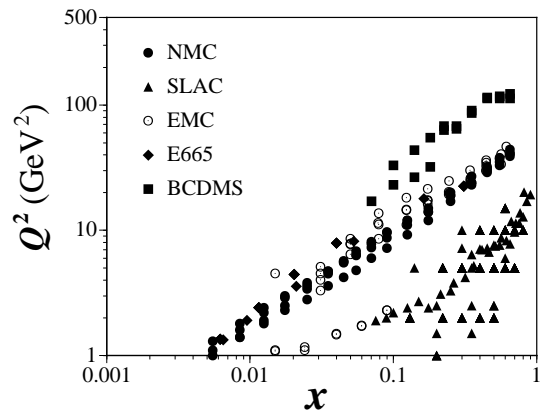}
\vspace{-0.2cm}
\caption{Kinematical range is shown by
         plotting $x$ and $Q^2$ points of the used data.}
\label{fig:xq2}
\end{figure}
We show the $x$ and $Q^2$ points of the employed data in
Fig. \ref{fig:xq2}. 
The SLAC data are restricted to the large $x$ and small $Q^2$ region. 
Because the SLAC data are taken for many nuclear species, their data
are very valuable for our analysis. However, they cannot address
the issue of shadowing due to the lack of small $x$ measurements.
The BCDMS data are also taken in the large $x$ region.
The difference from the SLAC measurements is that
the BCDMS data have large $Q^2$ values. There exist a large $Q^2$ gap
between the SLAC and BCDMS data sets, which may enable us
to investigate nuclear $Q^2$ evolution. 
On the other hand, the EMC, NMC, E665 data are almost in
the same kinematical range. Because these data include small
$x$ points, it is possible to investigate the shadowing
region as well as the medium-$x$ modification part.
However, the data have rather small $Q^2$ values
in a restricted $Q^2$ range at small $x$. It suggests that
it is difficult to determine the nuclear gluon distributions
from the scaling violation at small $x$.
In order to obtain the smaller $x$ or larger $Q^2$ data than
those in Fig. \ref{fig:xq2}, we should wait for a next generation
project such as HERA-eA \cite{ahera} or eRHIC \cite{erhic}.

\subsection{$\chi^2$ analysis}
\label{chi2}

Our analysis is done in the leading order (LO) of $\alpha_s$.
The structure function $F_2^A$ is expressed in a parton model as
\begin{equation}
F_2^A (x,Q^2) = \sum_q e_q^2 x [ q^A(x,Q^2) + \bar q^A(x,Q^2) ],
\end{equation}
where $e_q$ is the quark charge,
and $q^A$ ($\bar q^A$) is the quark (antiquark) distribution
in the nucleus $A$.
It is noteworthy to mention here that the structure functions 
and the parton distributions are defined by those per nucleon
throughout this paper.
Although it is now established that each antiquark
distribution is different in the nucleon \cite{skpr},
there is no data which could suggest flavor dependent antiquark
distributions in a nucleus \cite{skantiq,fnal-pro}. 
Therefore, we should inevitably assume flavor symmetric antiquark
distributions:
\begin{equation}
\bar u^A = \bar d^A = \bar s^A \equiv \bar q^A.
\end{equation}
Furthermore, the flavor number is taken three.
Then, the structure function becomes a summation of
valence-quark and antiquark distributions:
\begin{equation}
F_2^A (x,Q^2) = \frac{x}{9} [ 4 u_v^A (x,Q^2) + d_v^A (x,Q^2) 
                               + 12 \bar q^A (x,Q^2) ] .
\end{equation}
The gluon distribution $g^A(x,Q^2)$ is not explicitly contained
in the LO structure function; however, it contributes to $F_2^A$ through
$Q^2$ evolution which is described by coupled integro-differential
equations with $g^A (x,Q^2)$.
In this way, we need four types of distributions,
$u_v^A$, $d_v^A$, $\bar q ^A$, and $g^A$, for calculating
the structure function $F_2^A$.

Because the valence-quark distributions in a nucleus
are much different from the ones in the proton, we should
be careful in defining the weight function $w_i (x,A,Z)$.
If there {\it were} no nuclear modification, in other words,
if a nucleus were described simply by a collection of protons and
neutrons, the parton distributions in the nucleus $A$ are given by
\begin{align}
A f_i^A (x,Q^2)_{\text{no-mod}}
                & = Z f_i^p (x,Q^2) + N f_i^n (x,Q^2), \nonumber \\
                & \text{where} \ f_i = u_v, d_v, \bar q, \text{or} \ g.
\end{align}
The functions $f_i^p (x,Q^2)$ and $f_i^n (x,Q^2)$ are $i$-type
distributions in the proton and neutron, respectively.
Isospin symmetry is usually assumed for the parton distributions
in the proton and the neutron:
\begin{alignat}{2}
u_v^n & = d_v^p \equiv d_v, & \ \ \ d_v^n & = u_v^p \equiv u_v,
\nonumber \\
\bar q^n & = \bar q^p \equiv \bar q, & \ \ \  g^n & = g^p \equiv g.
\end{alignat}
Then, the nuclear parton distributions become
\begin{align}
u_v^A (x,Q^2)_{\text{no-mod}} & = \frac{Z u_v (x,Q^2) + N d_v (x,Q^2)}{A},
\nonumber \\
d_v^A (x,Q^2)_{\text{no-mod}} & = \frac{Z d_v (x,Q^2) + N u_v (x,Q^2)}{A},
\nonumber \\
\bar q^A (x,Q^2)_{\text{no-mod}} & = \bar q (x,Q^2),         \nonumber \\
g^A (x,Q^2)_{\text{no-mod}}      & = g (x,Q^2).
\label{eqn:pn}
\end{align}

As suggested by the $F_2^A/F_2^D$ measurements, 10$-$30\% modification
is expected for medium and large size nuclei. The modification from
the expressions in Eq. (\ref{eqn:pn}) should be expressed by the
functions $w_i (x,A,Z)$ at $Q_0^2$ as discussed in Sec. \ref{ax-depend}:
\begin{align}
u_v^A (x,Q_0^2) & = w_{u_v} (x,A,Z) \, \frac{Z u_v (x,Q_0^2) 
                                         + N d_v (x,Q_0^2)}{A},
\nonumber \\
d_v^A (x,Q_0^2) & = w_{d_v} (x,A,Z) \, \frac{Z d_v (x,Q_0^2) 
                                         + N u_v (x,Q_0^2)}{A},
\nonumber \\
\bar q^A (x,Q_0^2) & = w_{\bar q} (x,A,Z) \, \bar q (x,Q_0^2),
\nonumber \\
g^A (x,Q_0^2)      & = w_{g} (x,A,Z) \, g (x,Q_0^2).
\label{eqn:wpart}
\end{align}
We would like to take small $Q_0^2$ value 
in order to accommodate experimental data as many as possible.
On the other hand, because the Dokshitzer-Gribov-Lipatov-Altarelli-Parisi
(DGLAP) equations are used for the $Q^2$ evolution, $Q_0^2$ should be
large enough so that perturbative QCD can be applied. 
As a point which could compromise these conflicting requirements,
$Q_0^2=$1 GeV$^2$ is taken.

Next, a set of parton distributions in the nucleon is selected.
There are available CTEQ, GRV, and MRS distributions. 
However, analytical expressions are not given in the CTEQ paper
\cite{cteq5}, and the GRV expressions are provided at
small $Q^2$ (0.26 GeV$^2$) \cite{grv98}.
Therefore, we decided to use a LO set of MRST (Martin, Roberts,
Stirling, and Thorne) \cite{mrst}, which is conveniently defined
at $Q_0^2$=1 GeV$^2$. In this way, we use the central gluon version of
MRST-LO distributions with the scale parameter $\Lambda_{LO}=0.1741$ GeV
in our $\chi^2$ analysis. Consequently, obtained nuclear distributions
in Sec. \ref{results} become the MRST distributions in the limit
$A \rightarrow 1$.

Because each function $w_i (x,A,Z)$ still has four or five parameters,
it is necessary to reduce the total number as many as possible
in order to become an efficient analysis.
First, there are following obvious constraints (a), (b), and (c)
for the nuclear distributions.

\noindent
{\bf (a) Charge}

Nuclear charge has to be the atomic number $Z$.
It can be expressed in a parton model by considering an infinite
momentum frame for the nucleus. Let us consider elastic scattering
of a real photon with its momentum $|\vec q| \rightarrow 0$
from a nucleus. Because each parton is moving very fast,
we could use a parton picture for describing the process.
Then, the nuclear charge is given as
\begin{align}
Z & = \int dx \, A \left [   \frac{2}{3} (u^A - \bar u^A)
                    - \frac{1}{3} (d^A - \bar d^A) 
                    - \frac{1}{3} (s^A - \bar s^A) 
           \right ]
\nonumber \\
  & = \int dx \, \frac{A}{3} \left [ 2 \, u_v^A - d_v^A \right ].
\label{eqn:charge}
\end{align}
The second equation is obtained because there is no net
strangeness in an ordinary nucleus although $s^A (x,Q^2)$
could be different from $\bar s^A (x,Q^2)$. 
The valence distributions are defined by
$u_v^A \equiv u^A - \bar u^A$ and $d_v^A \equiv d^A - \bar d^A$
as usual.

\noindent
{\bf (b) Baryon number}

Baryon number of a nucleus is $A$, and it is expressed
in the parton model as
\begin{align}
A & = \int dx \, A \left [ \frac{1}{3} (u^A - \bar u^A)
                    + \frac{1}{3} (d^A - \bar d^A) 
                    + \frac{1}{3} (s^A - \bar s^A) 
           \right ]
\nonumber \\
  & = \int dx  \, \frac{A}{3} \left [  u_v^A + d_v^A  \right ].
\label{eqn:baryon}
\end{align}

\noindent
{\bf (c) Momentum}

Nuclear momentum is the addition of each parton contribution:
\begin{align}
A & = \int dx \, A \, x  \left [ u^A + \bar u^A
                    + d^A + \bar d^A
                    + s^A + \bar s^A + g^A \right ]
\nonumber \\
  & = \int dx \, A \, x 
            \left [  u_v^A + d_v^A  + 6 \, \bar q^A + g^A \right ].
\label{eqn:momentum}
\end{align}

If the weight functions are the quadratic functional type,
the distributions are expressed by the parameters,
$a_{u_v}$, $a_{d_v}$, $b_v$, $c_v$, $\beta_v$, 
$a_{\bar q}$, $b_{\bar q}$, $c_{\bar q}$, $\beta_{\bar q}$,
$a_g$, $b_g$, $c_g$, and $\beta_g$.
Here, the valence up- and down-quark parameters are assumed
to be the same except for $a_{u_v}$ and $a_{d_v}$
because both weight functions are expected to be similar.
However, at least one parameter should be different in order
to satisfy the conditions (a) and (b) simultaneously
since there are data for non-isoscalar nuclei with $Z \ne N$.
Among the parameters, three of them can be fixed by 
the conditions, (a), (b), and (c).
We use these three conditions for determining $a_{u_v}$,
$a_{d_v}$, and $a_g$.

There are still irrelevant parameters which could be removed
from the parametrization.
First, $\beta_{\bar q}$ and $\beta_g$ describe the functional
shape of the antiquark and gluon distributions at large $x$.
However, they are irrelevant in the sense that
both distributions are extremely small at large $x$,
for example $x\sim 0.8$. They are not expected to contribute
to $F_2^A$ significantly in this $x$ region. Furthermore,
the antiquark and gluon distributions themselves are
not determined at such large $x$ in the nucleon. 
In this way, we decided to fix the parameters at 
$\beta_{\bar q}=\beta_g=1$. We checked the sensitivity of
our $\chi^2$ analysis results to this choice. Even if 
$\beta_{\bar q}=\beta_g=0.5$ or 2 is taken, we found that
the obtained $\chi^2$ changed very little. This fact indicates
that $\chi^2$ is almost independent of these parameters.
There is another irrelevant parameter. 
The gluon parameters $b_g$ and $c_g$ determine the functional
form at medium and large $x$. However, the gluon distribution
does not contribute to $F_2^A$ directly, so that the detailed
$x$ dependence cannot be determined in the analysis. Therefore,
$b_g$ is fixed at $b_g=-2 c_g$ with the following consideration.
The $\chi^2$ analysis tends to favor negative $c_g$.
As far as $b_g$ is taken to be larger than $- c_g$,
the choice does not affect the $\chi^2$ to a considerable extent.
However, if $b_g$ is taken smaller than
$-c_g$, it could contradict with the condition
$w_i(x\rightarrow 1,A,Z)\rightarrow +\infty$
depending on the value of $a_g$.
In this way, there are seven free parameters
\begin{equation}
b_v, \ \ c_v, \ \ \beta_v, \ \ 
a_{\bar q}, \ \ b_{\bar q}, \ \ c_{\bar q}, \ \  c_g,
\label{eqn:para-q}
\end{equation}
in the quadratic fit. 
There are additional parameters in the cubic fit. However,
as far as the gluon distribution is concerned, we use the
same quadratic form even in the cubic type analysis.
The structure function $F_2$ is rather insensitive to the gluon
distribution, especially in the LO analysis. It does not make
much sense to introduce an additional parameter for the gluon in
the $\chi^2$ analysis without the data which could restrict
the gluon distribution. In this way, the actual parameters are
\begin{equation}
b_v, \ \ c_v, \ \ d_v, \ \ \beta_v, \ \ 
a_{\bar q}, \ \ b_{\bar q}, \ \ c_{\bar q}, \ \ d_{\bar q}, \ \ 
c_g,
\label{eqn:para-c}
\end{equation}
in the cubic fit.

In the theoretical calculations, the nuclei are assumed as
$^4$He, $^7$Li, $^9$Be, $^{12}$C, $^{14}$N, $^{27}$Al, $^{40}$Ca,
$^{56}$Fe, $^{63}$Cu, $^{107}$Ag, $^{118}$Sn,
$^{131}$Xe, $^{197}$Au, and $^{208}$Pb. 
The initial nuclear distributions are provided at $Q^2=1$ GeV$^2$
with the parameters in Eq. (\ref{eqn:para-q}) or (\ref{eqn:para-c}).
They are evolved to the experimental $Q^2$ points by the DGLAP
evolution equations. Then, obtained structure-function ratios
$R_{F_2}^A$ are compared with the experimental values for calculating
\begin{equation}
\chi^2 = \sum_j \frac{(R_{F_2,j}^{A,data}-R_{F_2,j}^{A,theo})^2}
                     {(\sigma_j^{data})^2},
\label{eqn:chi2}
\end{equation}
where the experimental error is given by
systematic and statistical errors as
$(\sigma_j^{data})^2 = (\sigma_j^{sys})^2 + (\sigma_j^{stat})^2$.
Although the deuteron structure function is sometimes assumed the same
as the one for the nucleon \cite{saga,helsinki}, the deuteron
modification is also taken into account in our analysis simply by
setting $A=2$.
With these preparations together with the CERN subroutine
{\sc Minuit} \cite{minuit}, the optimum parameter set is obtained 
by minimizing $\chi^2$.

\section{Results}
\label{results}
\setcounter{equation}{0}

Our analysis results are explained in Sec. \ref{comp} in comparison
with the used experimental data. Then, obtained optimum nuclear
parton distributions are discussed in Sec. \ref{dist}.

\subsection{Comparison with data}
\label{comp}

Analysis results are shown for both quadratic and cubic types.
A minimal functional form is the quadratic type 
according to the discussion in Sec. \ref{x-depend}.
It is minimal in the sense that the major features of the measured
ratios $R^A_{F_2}$ could be described in the whole $x$ region including
increase at large $x \sim 0.9$, depletion at medium at $x \sim 0.6$, 
antishadowing at $x \sim 0.15$, and shadowing at small $x<0.05$
with a minimum number of parameters. However, there are some restrictions
on the distribution shape, most significantly on the valence-quark
distributions. In the cubic type analysis, the shape becomes more
flexible due to the additional parameters.
Because of the new freedoms, obtained $\chi_{min}^2$ should be smaller 
for the cubic type, however, by sacrificing the computation time. 

\begin{figure}[b!]
\includegraphics[width=0.41\textwidth]{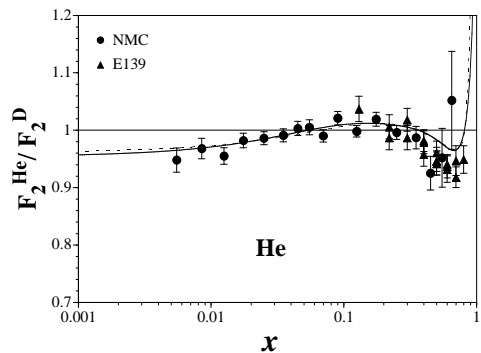}
\vspace{-0.3cm}
\caption{Fitting results are compared with the helium data.
         The dashed and solid curves are for
         the quadratic and cubic types, respectively,
         at $Q^2=5$ GeV$^2$, whereas the data are taken
         at various $Q^2$ points.}
\label{fig:he}
\end{figure}

\begin{figure}[t!]
\includegraphics[width=0.41\textwidth]{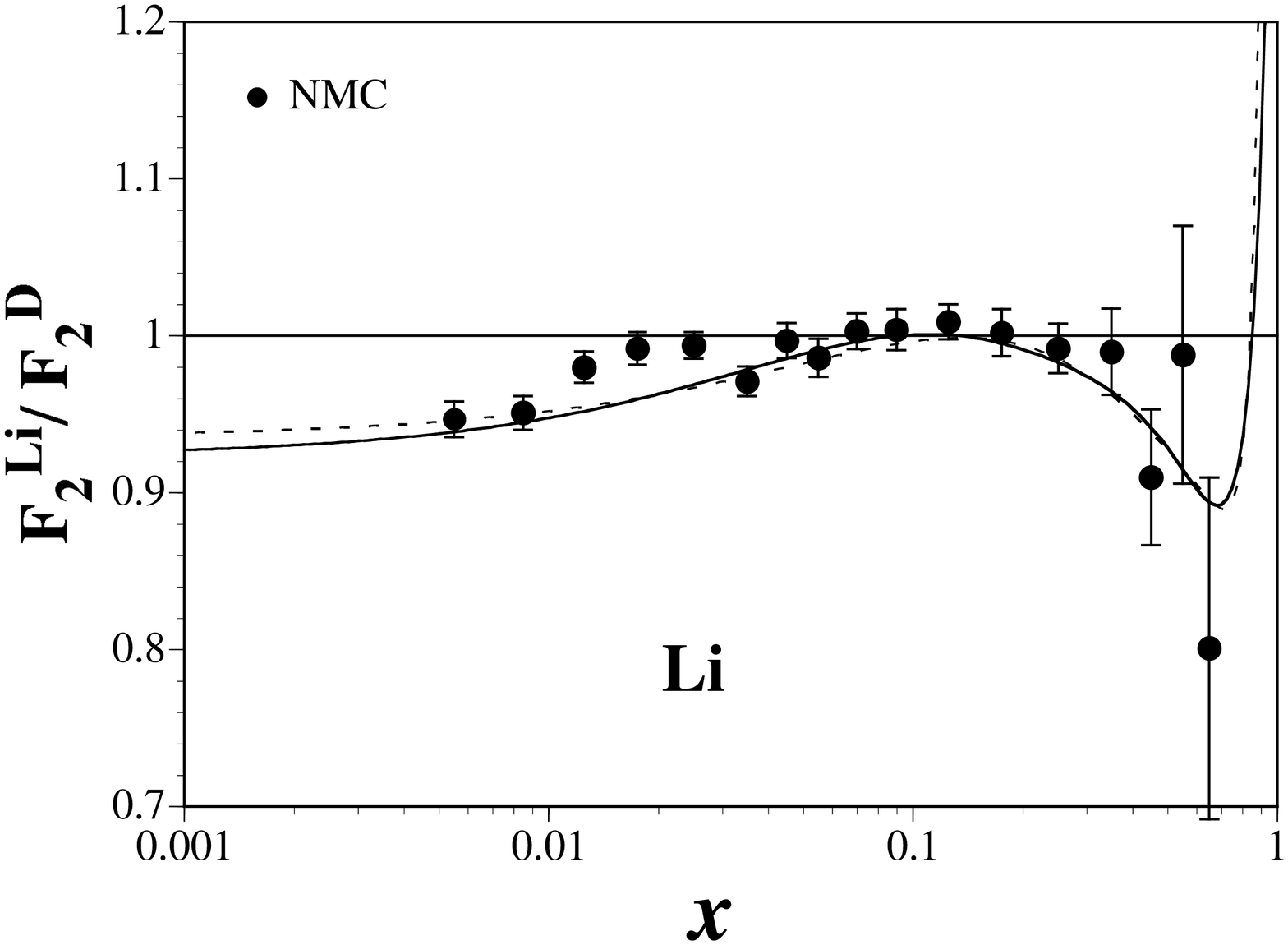}
\vspace{-0.7cm}
\caption{Comparison with the lithium data.}
\label{fig:li}
\vspace{0.3cm}
\includegraphics[width=0.41\textwidth]{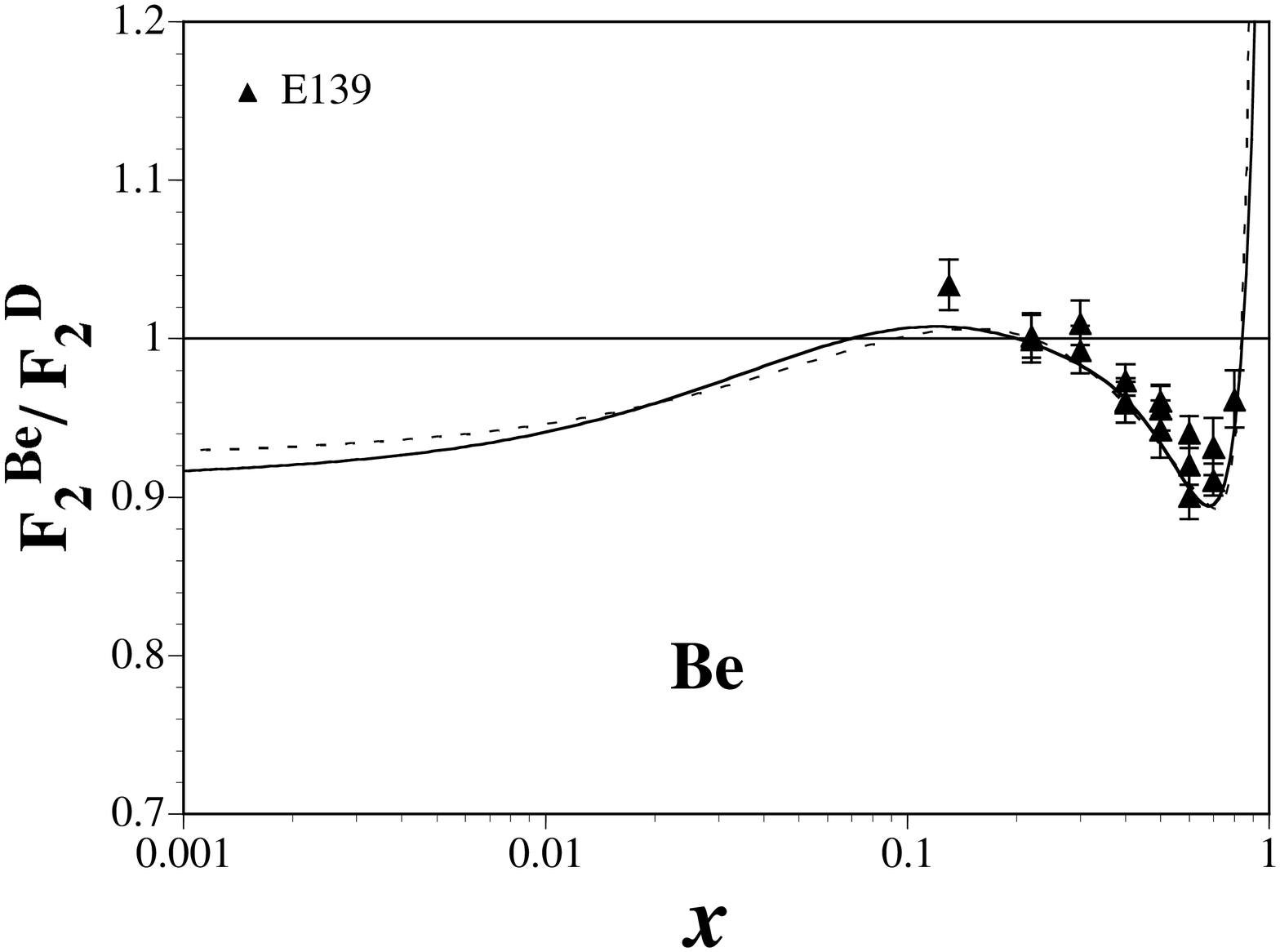}
\vspace{-0.7cm}
\caption{Comparison with the beryllium data.}
\label{fig:be}
\vspace{0.3cm}
\includegraphics[width=0.41\textwidth]{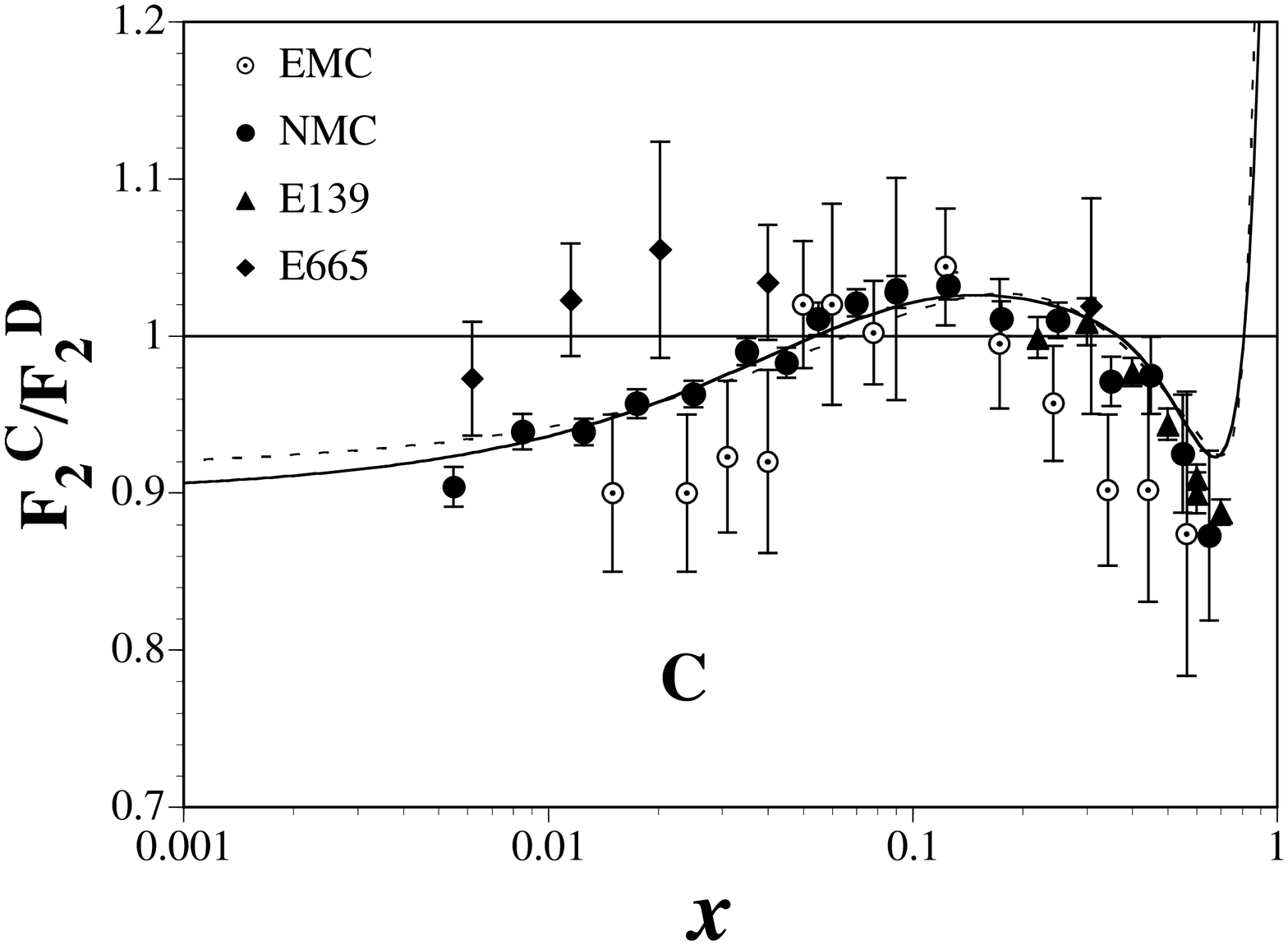}
\vspace{-0.7cm}
\caption{Comparison with the carbon data.}
\label{fig:c}
\vspace{0.3cm}
\includegraphics[width=0.41\textwidth]{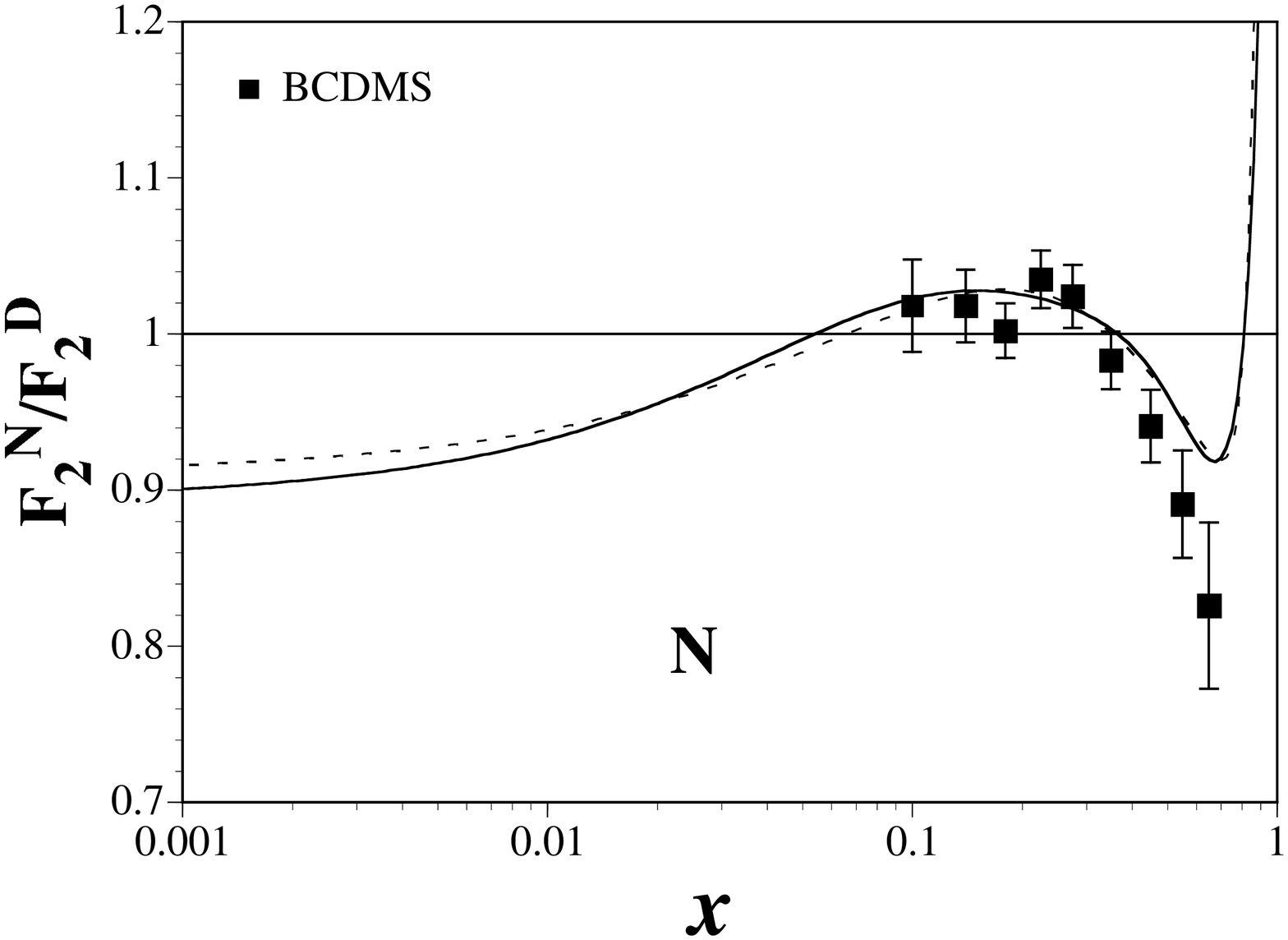}
\vspace{-0.7cm}
\caption{Comparison with the nitrogen data.}
\label{fig:n}
\end{figure}

\begin{figure}[t!]
\includegraphics[width=0.41\textwidth]{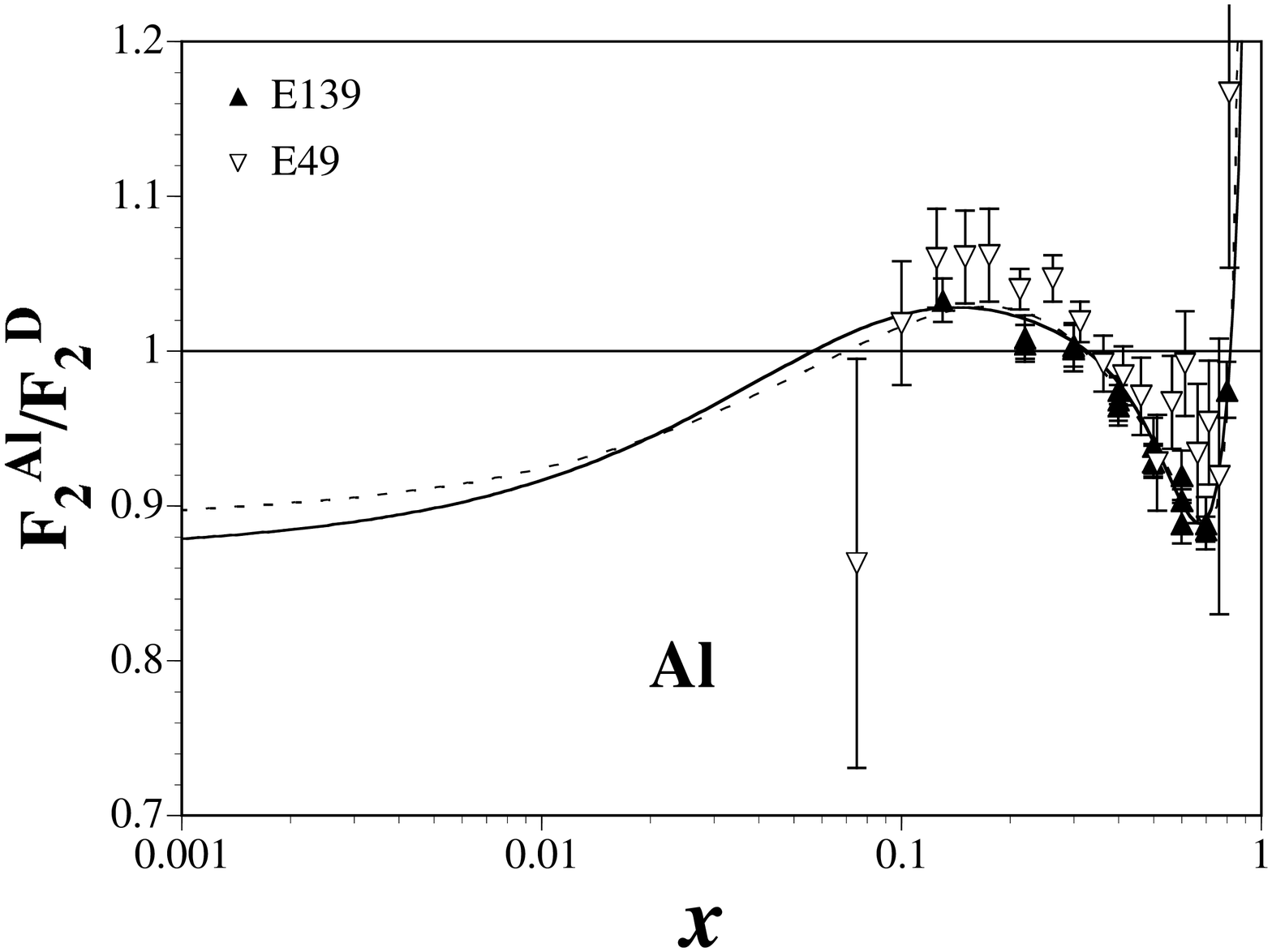}
\vspace{-0.3cm}
\caption{Comparison with the aluminum data.}
\label{fig:al}
\vspace{1.2cm}
\includegraphics[width=0.41\textwidth]{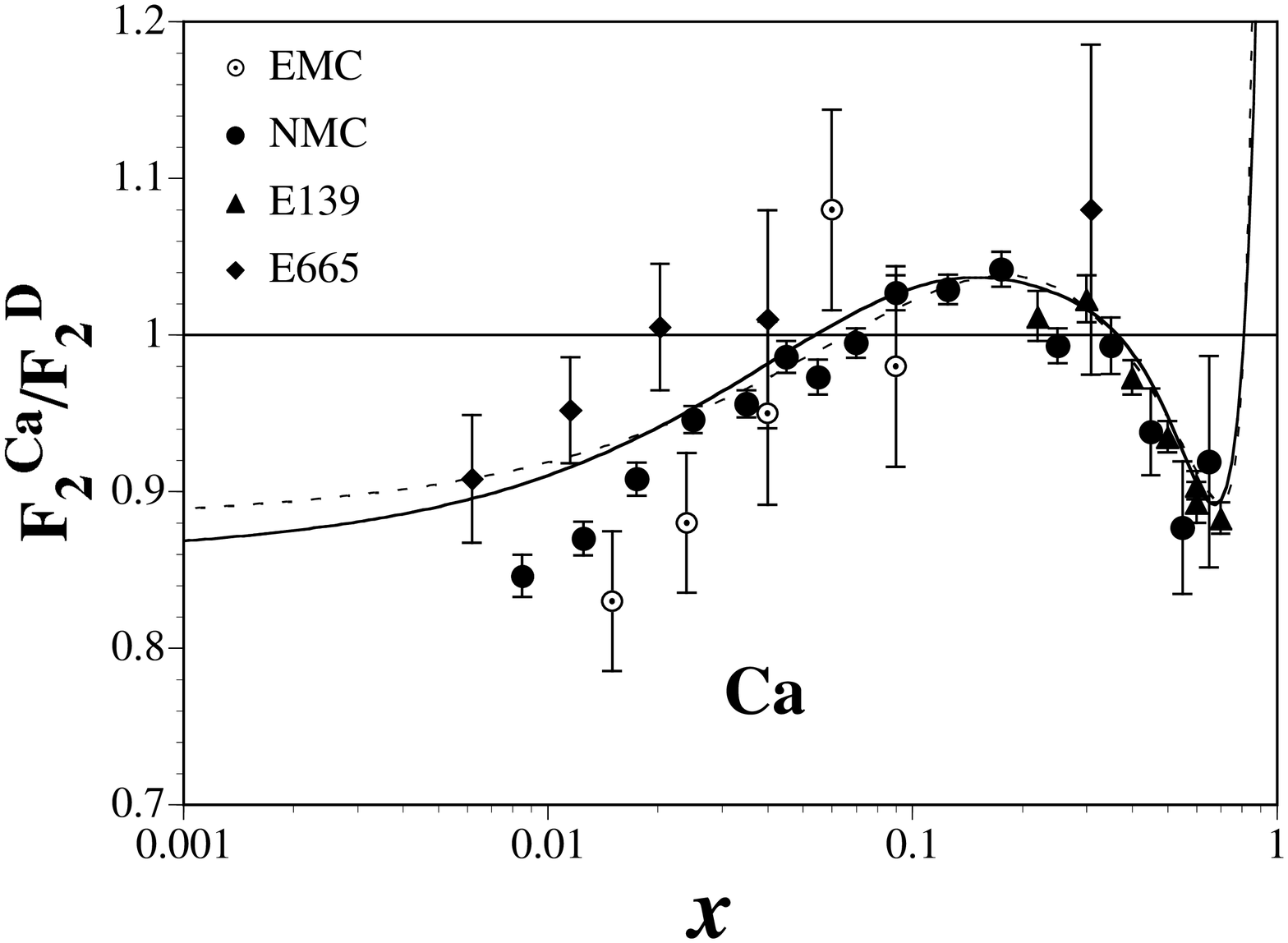}
\vspace{-0.3cm}
\caption{Comparison with the calcium data.}
\label{fig:ca}
\vspace{1.2cm}
\includegraphics[width=0.41\textwidth]{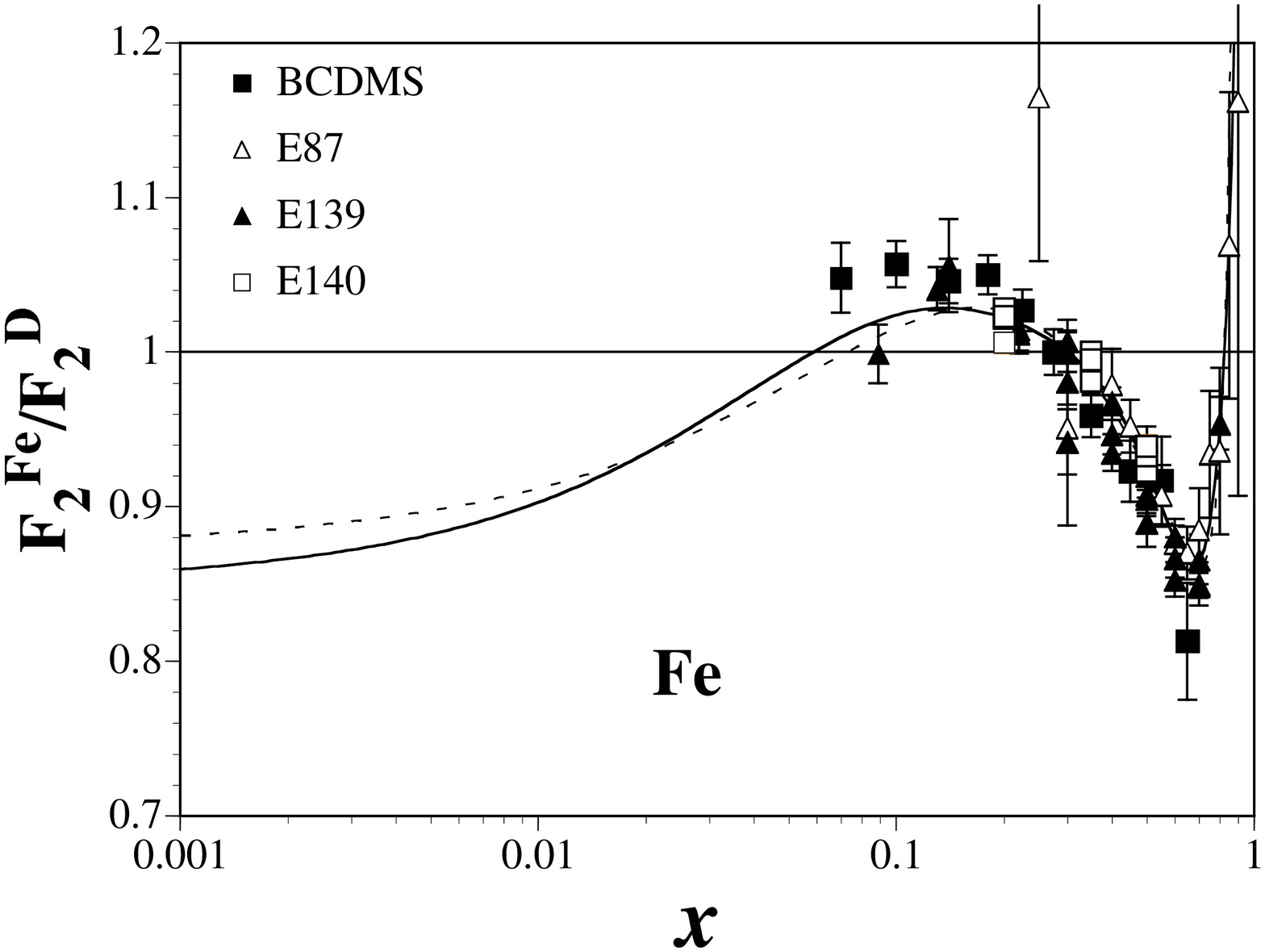}
\vspace{-0.3cm}
\caption{Comparison with the iron data.}
\label{fig:fe}
\end{figure}

\begin{figure}[t!]
\includegraphics[width=0.41\textwidth]{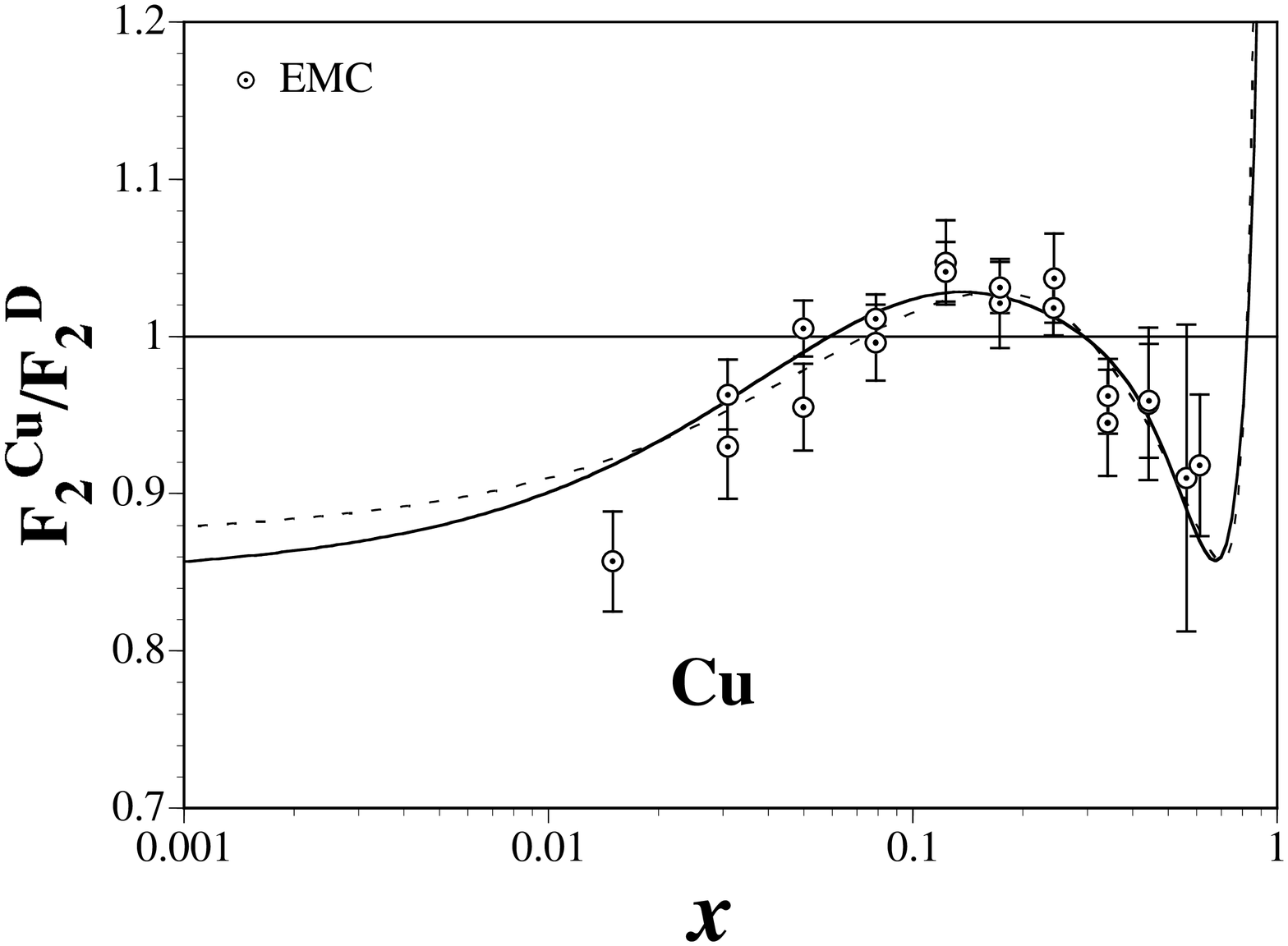}
\vspace{-0.3cm}
\caption{Comparison with the copper data.}
\label{fig:cu}
\vspace{1.3cm}
\includegraphics[width=0.41\textwidth]{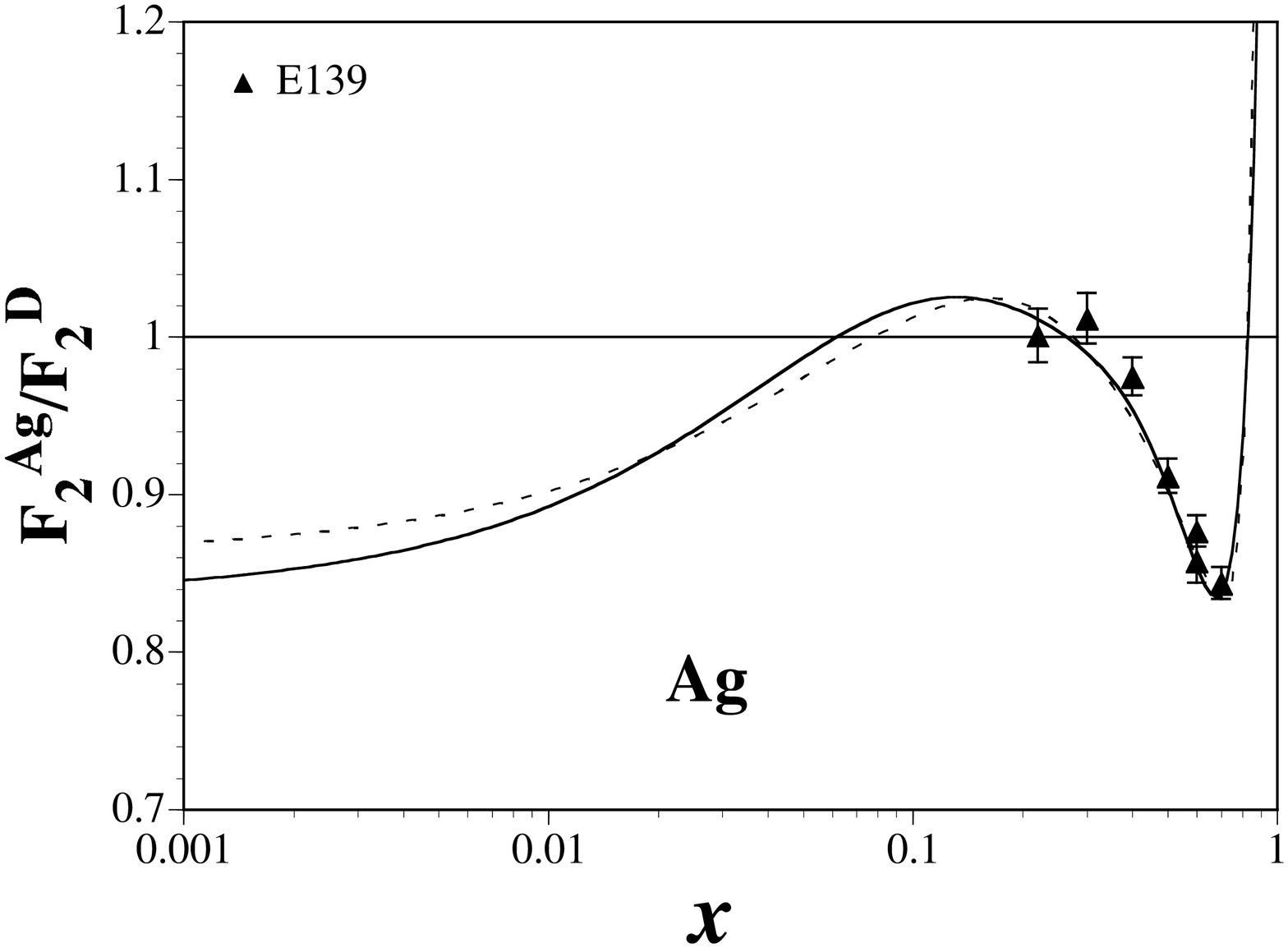}
\vspace{-0.3cm}
\caption{Comparison with the silver data.}
\label{fig:ag}
\vspace{1.3cm}
\includegraphics[width=0.41\textwidth]{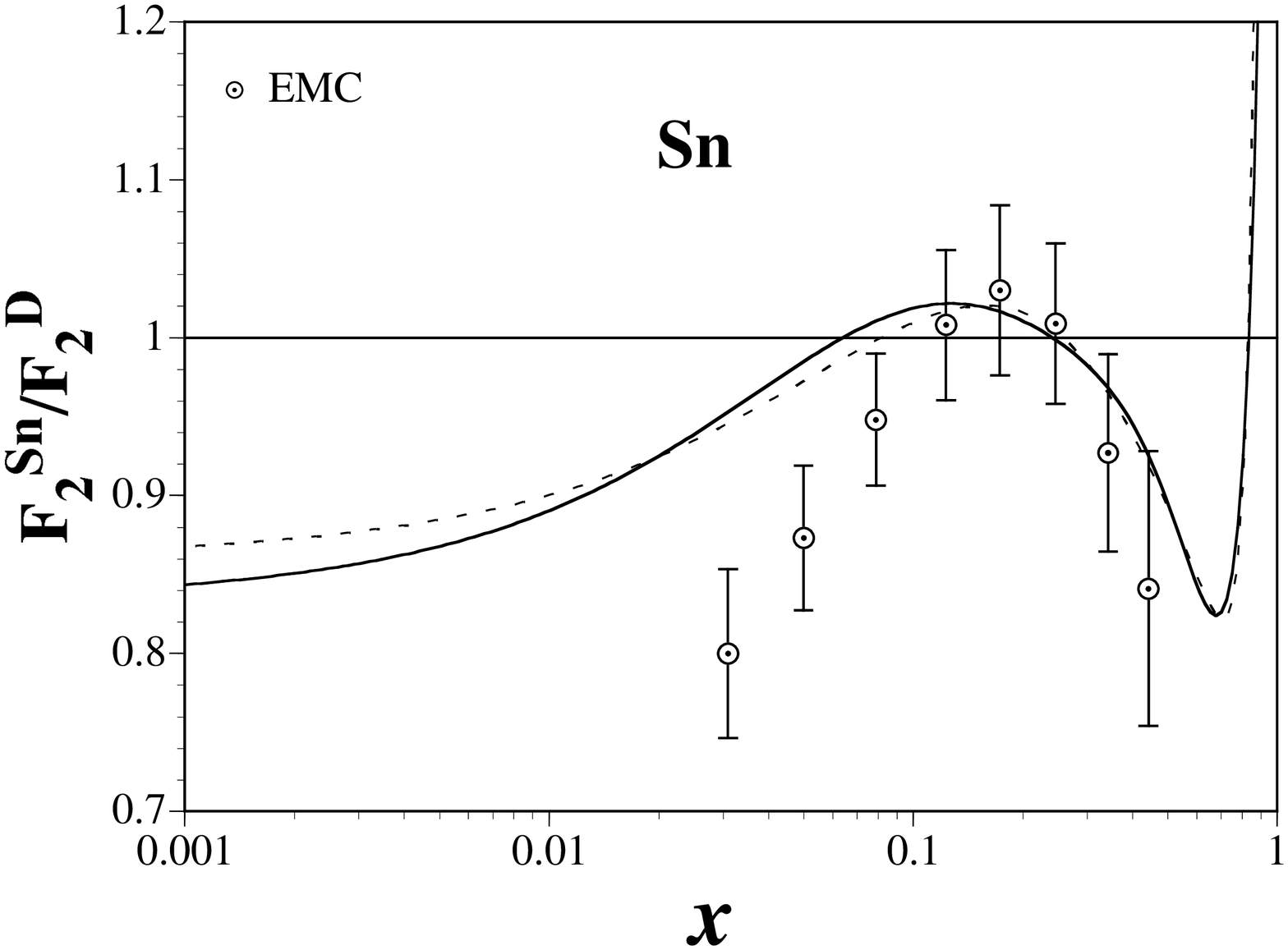}
\vspace{-0.3cm}
\caption{Comparison with the tin data.}
\label{fig:sn}
\end{figure}

\begin{figure}[t!]
\includegraphics[width=0.41\textwidth]{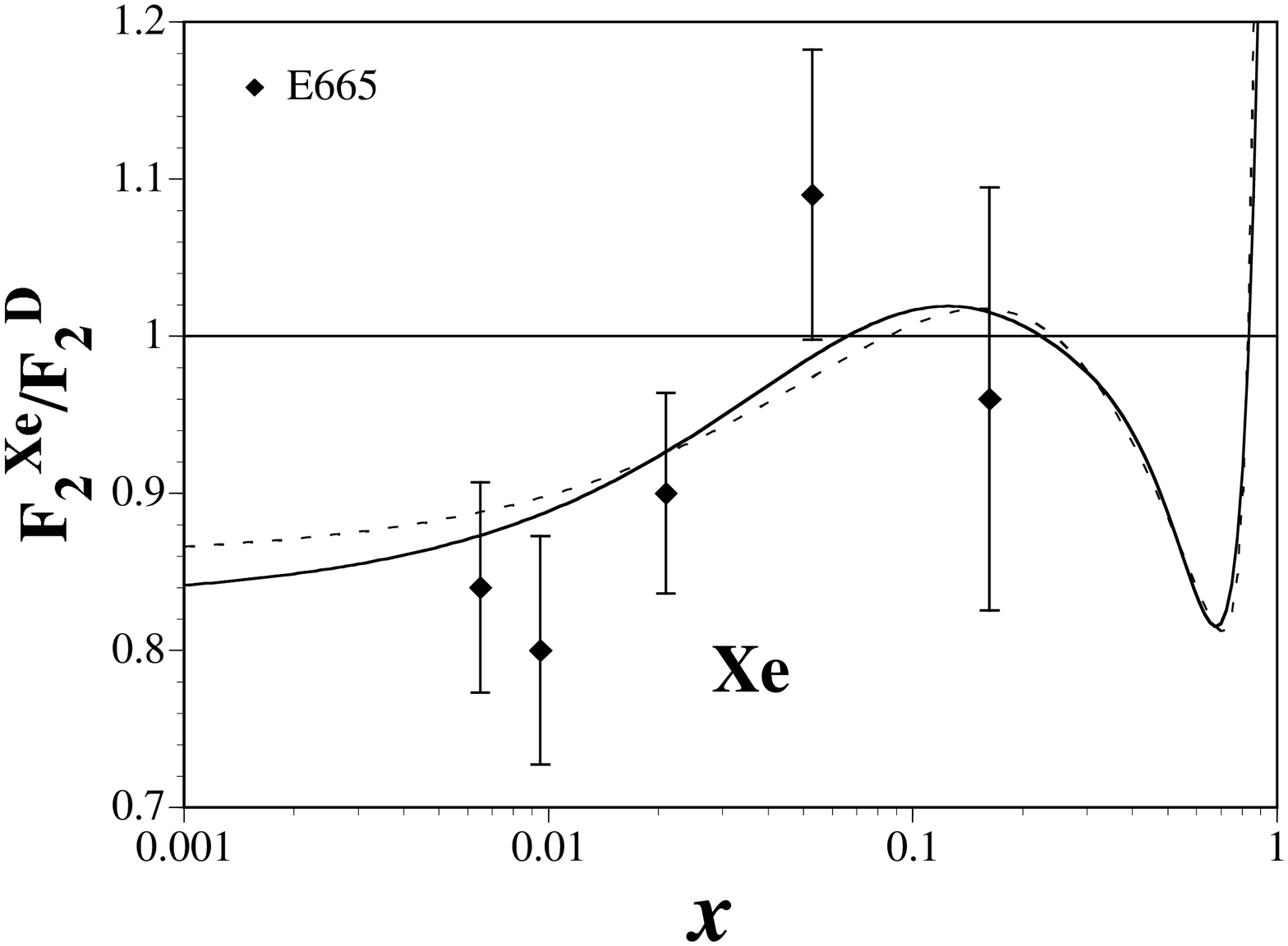}
\vspace{-0.3cm}
\caption{Comparison with the xenon data.}
\label{fig:xe}
\vspace{1.3cm}
\includegraphics[width=0.41\textwidth]{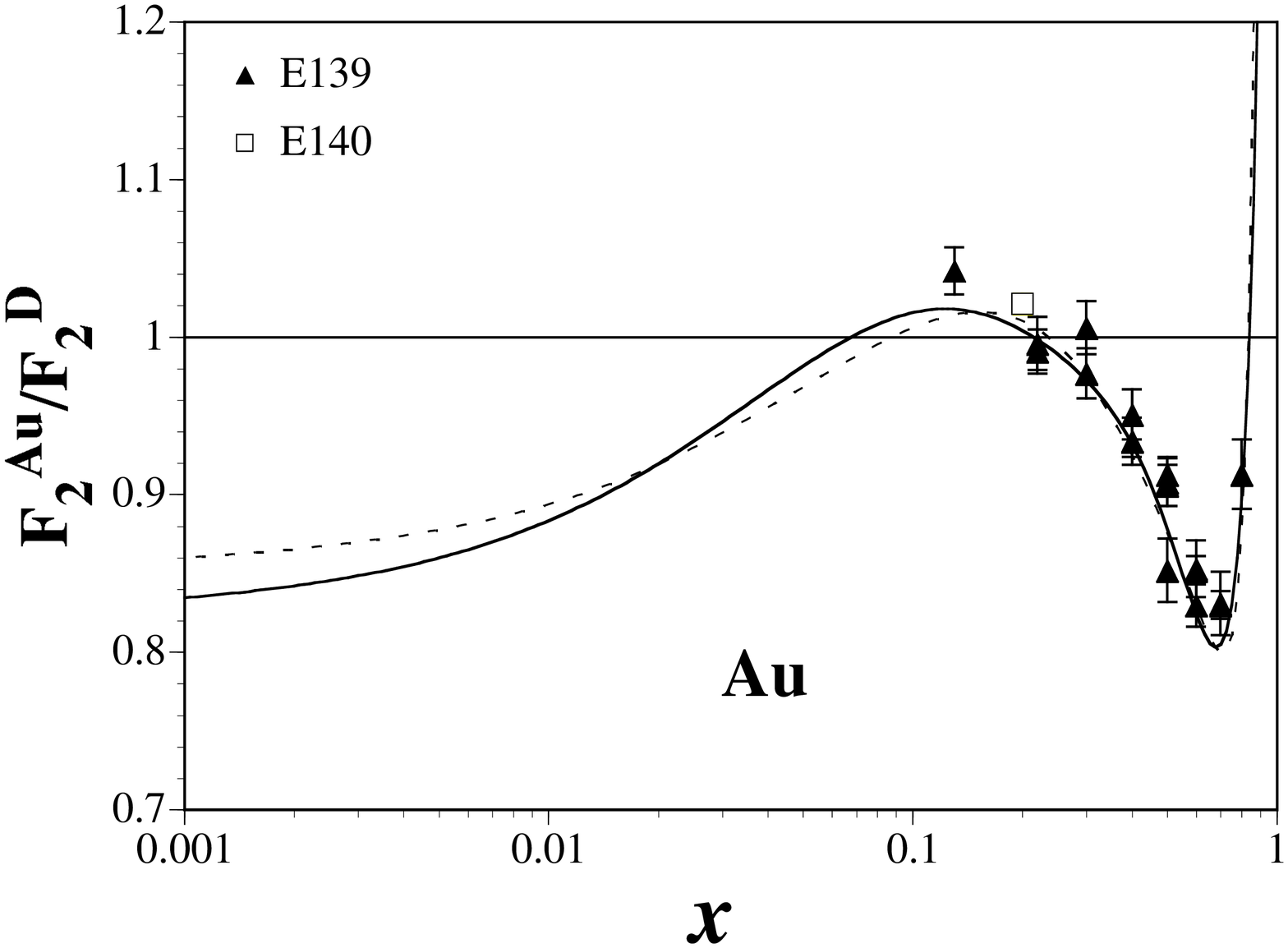}
\vspace{-0.3cm}
\caption{Comparison with the gold data.}
\label{fig:au}
\vspace{1.3cm}
\includegraphics[width=0.41\textwidth]{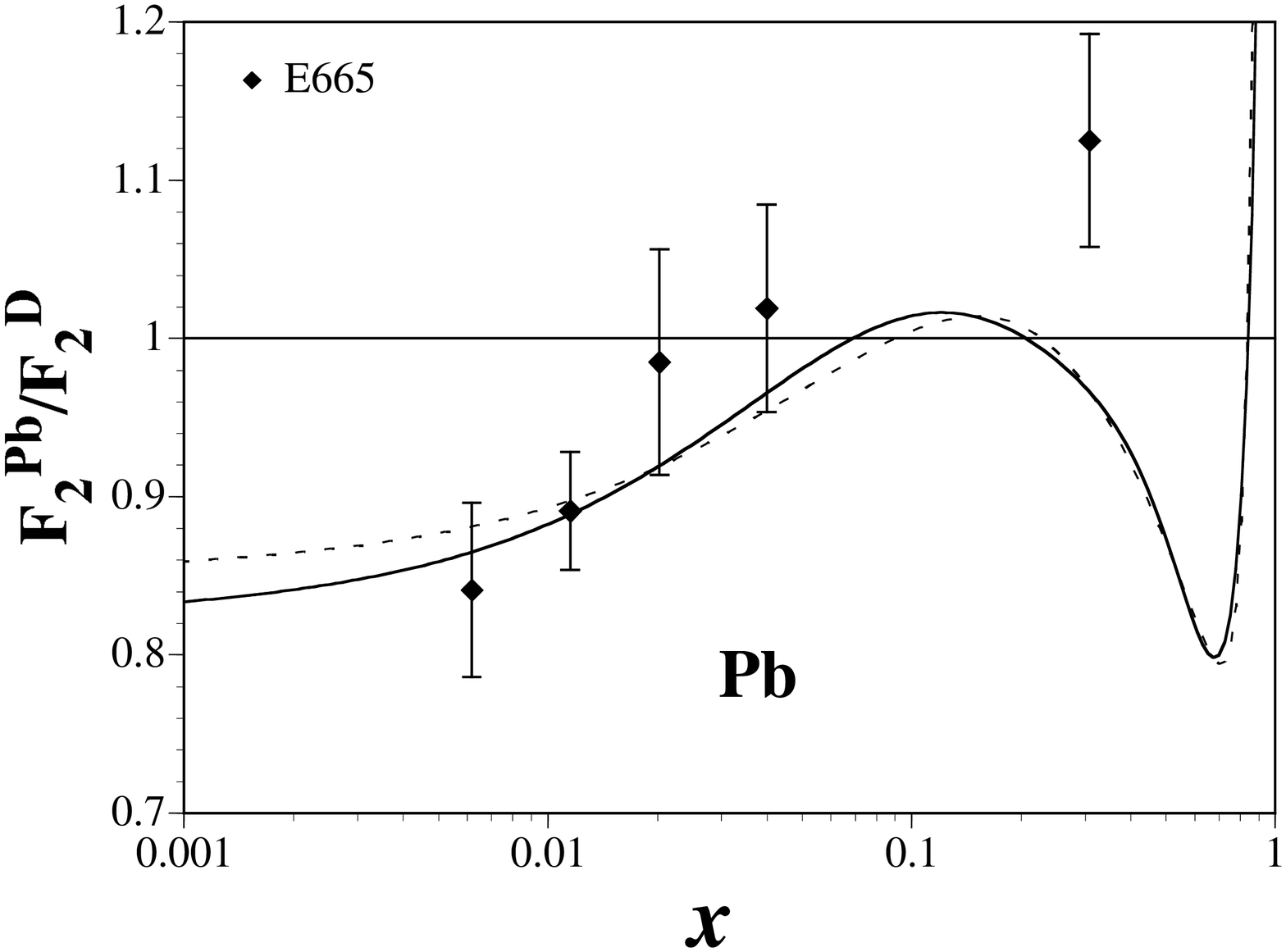}
\vspace{-0.3cm}
\caption{Comparison with the lead data.}
\label{fig:pb}
\end{figure}

Fitting results are compared with experimental data for
all the used nuclear data in Figs. \ref{fig:he}$-$\ref{fig:pb}.
Obtained optimum parton distributions are used for calculating
the curves at $Q^2$=5 GeV$^2$ in these figure.
The dashed and solid curves indicate the ratios
in the quadratic and cubic analyses, respectively.
The experimental data are taken at various $Q^2$ points
as shown in Fig. \ref{fig:xq2}, so that
the data cannot be compared directly with the curves;
nevertheless we can see a general tendency.

Obtained minimum $\chi^2$ values are $\chi_{min}^2$=583.7 and 546.6
in the quadratic and cubic analyses, respectively, for
the 309 total data points. Because the $\chi^2$ per degrees of
freedom is given by $\chi_{min}^2/d.o.f.$=1.93 (quadratic)
and 1.82 (cubic), they may not seem to be excellent fits. 
However, it is partly due to scattered experimental data
as it is obvious, for example, from Fig. \ref{fig:ca}.
The data from different experimental groups are scattered particularly
in the small $x$ region, and they contribute to $\chi^2$
significantly. Therefore, a slightly large $\chi_{min}^2$
($\chi_{min}^2/d.o.f.>1$) is unavoidable whatever the analysis method is.

From the figures, we find that the experimental shadowing
at small $x<0.05$ and anti-shadowing at $x\sim 0.15$
are generally well reproduced by the analysis.
There are slight deviations from the data at medium $x$
for carbon and nitrogen. However, if we try to explain the carbon and
nitrogen data at medium $x$, we overestimate the depletion for
the beryllium, silver, and gold as obvious from Figs. \ref{fig:be},
\ref{fig:ag}, and \ref{fig:au}. The fit is also not excellent
for the helium at medium $x$. The reason could be that the helium
nucleus is an exceptional tightly-bound system
which cannot be explained by the simple $1/A^{1/3}$ behavior.

There are typical differences between 
the quadratic and cubic curves in Figs. \ref{fig:he}$-$\ref{fig:pb}. 
They are different in the small $x$ region, where there are not
so many experimental data. The quadratic curves are above the cubic
ones at small $x$ ($0.001<x<0.01$), but they are below
in the region $0.03< x < 0.14$. Both curves agree well in
the larger $x$ region, where there are many experimental data.
From these figures, we find that both analyses results are similar
except for the minor differences in the small $x$ region.

\vspace{0.3cm}
\begin{table}[h]
\caption{Each $\chi^2$ contribution.}
\label{tab:chi2}
\begin{ruledtabular}
\begin{tabular*}{\hsize}
{c@{\extracolsep{0ptplus1fil}}c@{\extracolsep{0ptplus1fil}}c
@{\extracolsep{0ptplus1fil}}c}
nucleus & \# of data & $\chi^2$ (quad.)  & $\chi^2$ (cubic)  \\
\colrule
He    &  35  &  \ 55.6   &  \     54.5  \\
Li    &  17  &  \ 45.6   &  \     49.2  \\ 
Be    &  17  &  \ 39.7   &  \     38.4  \\
C     &  43  &  \ 97.8   &  \     88.2  \\
N     & \ 9  &  \ 10.5   &  \     10.4  \\
Al    &  35  &  \ 38.8   &  \     41.4  \\
Ca    &  33  &  \ 72.3   &  \     69.7  \\
Fe    &  57  &   115.7   &  \     92.7  \\
Cu    &  19  &  \ 13.7   &  \     13.6  \\
Ag    & \ 7  &  \ 12.7   &  \     11.5  \\
Sn    & \ 8  &  \ 14.8   &  \     17.7  \\
Xe    & \ 5  & \ \ 3.2   &  \  \ \ 2.4  \\
Au    &  19  &  \ 55.5   &  \     49.2  \\
Pb    & \ 5  & \ \ 7.9   &  \  \ \ 7.6  \\	
\colrule 
total & 309  &   583.7   &  \    546.6  \\
\end{tabular*}
\end{ruledtabular}
\end{table}
\vspace{1.0cm}

Each $\chi^2$ contribution is listed in Table \ref{tab:chi2}.
We notice that the $\chi^2$ values per data are especially larger
for the carbon, calcium, and gold nuclei than the average.
Because the NMC errors are very small, slight deviations
from the NMC data produce large $\chi^2$ values.
For example, the calcium data at $x=0.25$ have 
peculiar behavior which cannot be reproduced by a smooth
$x$ dependent function, yet they have small errors which
contribute significantly to the total $\chi^2$.

From the cubic $\chi^2$ values in Table \ref{tab:chi2}
in comparison with the quadratic ones, 
we find significant $\chi^2$ improvements in carbon, calcium, iron,
and gold, however, by sacrificing the $\chi^2$ values for lithium,
aluminum, and tin. The additional degrees of freedom make it possible to
adjust the distribution shapes to the data. There is a 20\% $\chi^2$
improvement in the iron from the quadratic analysis to the cubic one;
however, the difference is not very clear in Fig. \ref{fig:fe}
within the region where the data exist, except for the $x\sim 0.1$ region. 

From Figs. \ref{fig:he}$-$\ref{fig:pb} and Table \ref{tab:chi2},
we find that the analyses with the quadratic and cubic functional types
are both successful for reproducing the major experimental properties.
It is obvious from the $\chi^2$ comparison that the cubic results
are better. However, because the $\chi^2$ improvement is not very
large and both curves look similar in Figs. \ref{fig:he}$-$\ref{fig:pb},
both results could be taken as possible nuclear distributions.

\subsection{Obtained parton distributions}
\label{dist}

\begin{figure}[b!]
\vspace{0.5cm}
\includegraphics[width=0.31\textwidth]{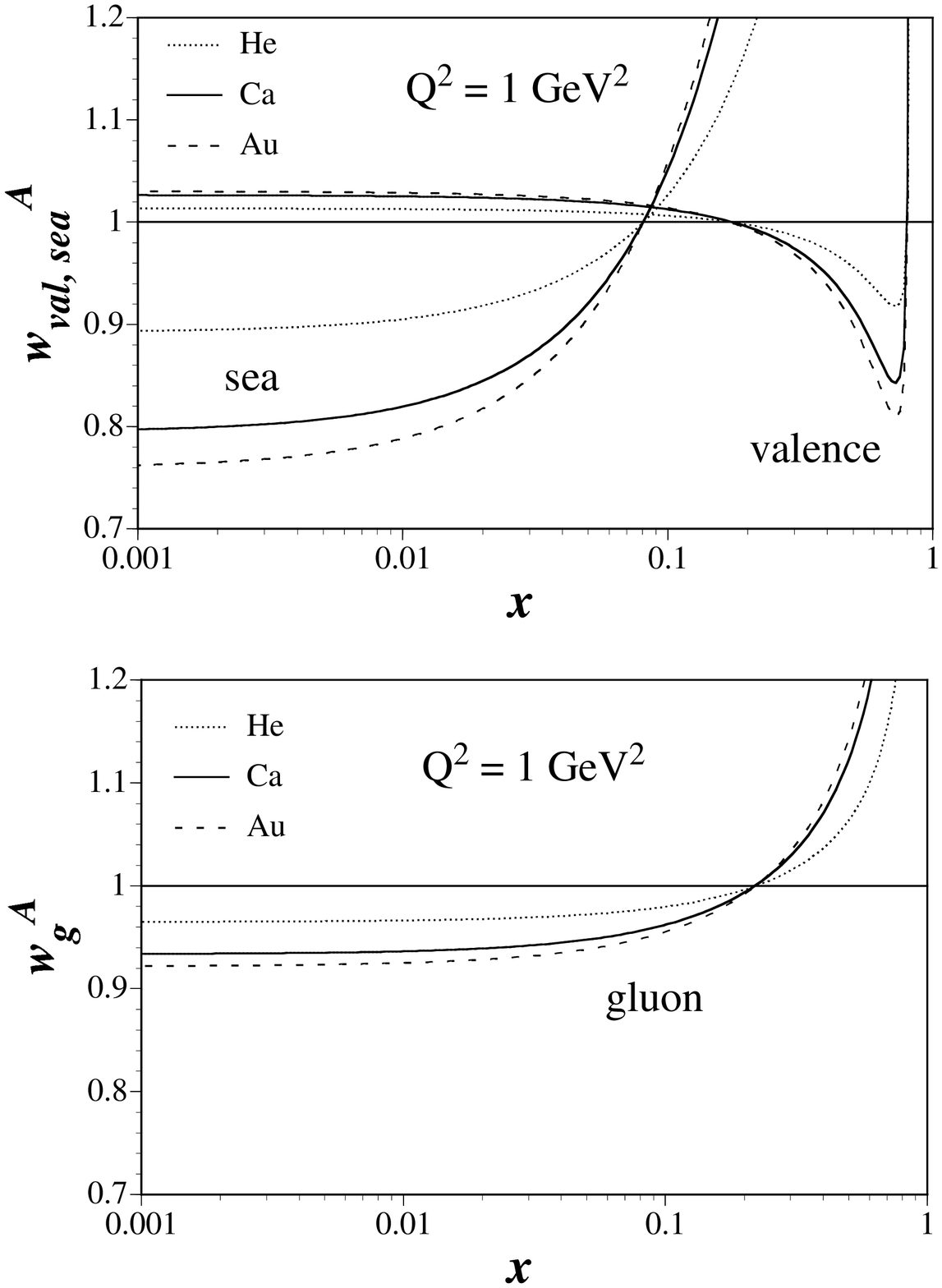}
\vspace{-0.8cm}
\caption{Obtained weight functions for the helium, calcium, and gold
         nuclei in the quadratic analysis.
         Only the valence up-quark functions are shown
         as the valence distributions.}
\label{fig:q-wxa}
\vspace{1.0cm}
\includegraphics[width=0.31\textwidth]{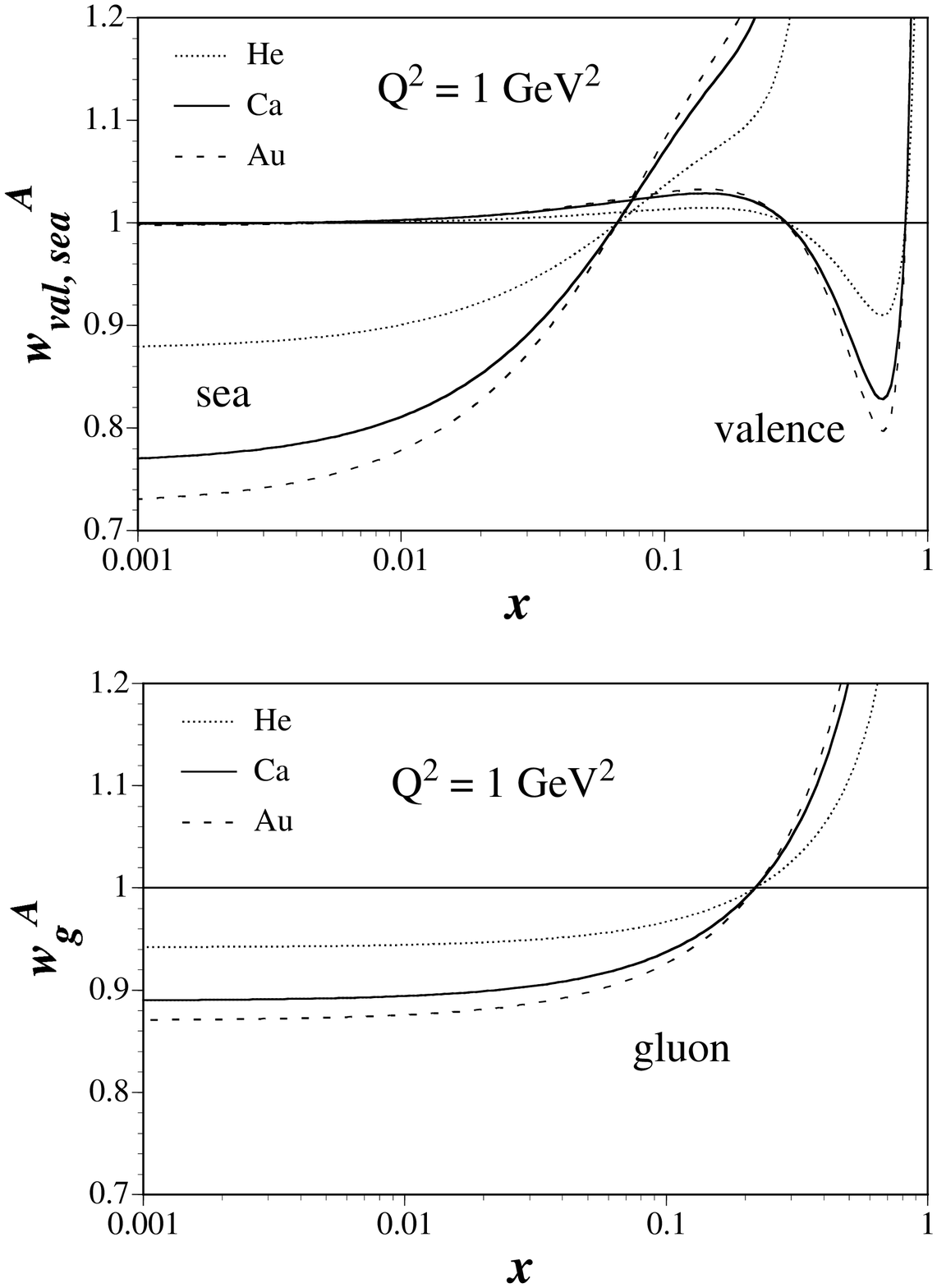}
\vspace{-0.8cm}
\caption{Obtained weight functions for the helium, calcium, and gold
         nuclei in the cubic analysis.
         Only the valence up-quark functions are shown
         as the valence distributions.}
\label{fig:c-wxa}
\end{figure}

Obtained optimum weight functions in the helium, calcium, and gold
nuclei are shown in Figs. \ref{fig:q-wxa} and \ref{fig:c-wxa}
for the quadratic and cubic analyses, respectively.
Because valence up- and down-quark functions are the same for
isoscalar nuclei and they are very similar in other nuclei,
only the valence up-quark functions are shown in these figures.

First, the quadratic results are explained. The valence-quark distributions
have depletion at medium $x$ because they should explain the modification
of the ratios $R_{F_2}^A$ at $x \sim 0.6$. Because of the assumed quadratic
functional form with the baryon-number constraint, the valence
distributions show antishadowing property at small $x$. 
It indicates 2.6\% antishadowing for the calcium nucleus
at $x =0.001$. It is noteworthy to reiterate that
this quadratic type could not allow a shadowing property for
the valence-quark distributions, so that it does not agree with
a shadowing prediction, for example, in Ref. \cite{fsl}
although it could agree with a parton-model analysis of
Refs. \cite{saga,F3A} and also with the one in Ref. \cite{kulagin}.
Next, the antiquark distributions should explain the shadowing
of $F_2$ at small $x$, so that they
also have shadowing property at $x \lesssim 0.07$. 
The antiquark shadowing is about 20\% for the calcium.
The antiquark weight functions increase as $x$ becomes larger.
They cross the line $w_{\bar q}=1$ at $x \sim 0.08$ and continue
to increase. The gluon weight functions have similar property
to the antiquark functions except that
the shadowing is smaller (7\% for Ca) and that the crossing point
is slightly larger ($x \sim 0.2$). The similar functional form
is obtained partly due to the momentum-conservation constraint. 
The gluon distribution does not contribute significantly 
in the LO analysis; however, our analysis tends to rule out
the gluon antishadowing at small $x$. 

We performed the cubic analysis for getting better agreement
with the data by the additional adjustable parameters. However,
as shown in Fig. \ref{fig:c-wxa}, the obtained functions are
similar to those in Fig. \ref{fig:q-wxa} except for the valence
functions at small $x$.
Now, the valence distributions have freedom to
have shadowing or antishadowing property at small $x$. 
In fact, the results show shadowing for the valence distributions
at very small $x$; however, the magnitude is fairly small,
0.1\% at $x$=0.001.
The bump of $R_{F_2}^A$ in the region, $0.1<x<0.2$, is explained mainly
by the antiquark distributions. This situation is very different,
for example, from the picture in Ref. \cite{fsl},
where the bump is explained mostly by the valence distributions.
Of course, the $F_2$ data are not enough to separate the valence
and sea distributions, so that we should wait for future experimental
activities, especially at a neutrino factory \cite{sknu},
for precise information.

The antiquark weight functions in the cubic analysis are similar to
the ones in the quadratic case. We expected a possibility of
much wild behavior. For example, it has shadowing
at small $x$, antishadowing at $x \sim 0.15$, depletion at medium $x$,
and rise at large $x$. However, even if the input antiquark
distributions are given by this functional type
in our $\chi^2$ analysis, they converge to
the functions in Fig. \ref{fig:c-wxa}. 

As mentioned in the previous section, the gluon distributions
are assumed to be the quadratic functional form even
in the ``cubic" fit. Therefore, the obtained functions are very
similar to the ones in Fig. \ref{fig:q-wxa}.
We also tried the cubic type for the gluon. 
The additional factor $d_g x^3$ controls the behavior 
at large $x$. However, the medium and large $x$ behavior 
is almost irrelevant for the gluon, especially in the LO analysis
of $F_2$. Therefore, it is very difficult to control the gluon
parameters in a meaningful way within the $\chi^2$ analysis.
For example, the $\chi^2$ fit could produce an unphysical
negative gluon distribution at large $x$ if loose bounds are
given for the parameters.
It is almost meaningless to introduce the additional freedom
at large $x$ without the data which could restrict
the gluon distributions themselves. This is the reason
why we decided to have the quadratic gluon distributions even
in the ``cubic" analysis.

We find in Figs. \ref{fig:q-wxa} and \ref{fig:c-wxa} that
the variations from the calcium ($A=40$) to gold ($A=197$)
are small. Therefore, the obtained parton distributions
could be extrapolated into the distributions in
a nucleus with a larger mass number ($A>208$),
which is outside the analyzed nuclei in this paper.

\begin{figure}[b!]
\vspace{0.0cm}
\includegraphics[width=0.40\textwidth]{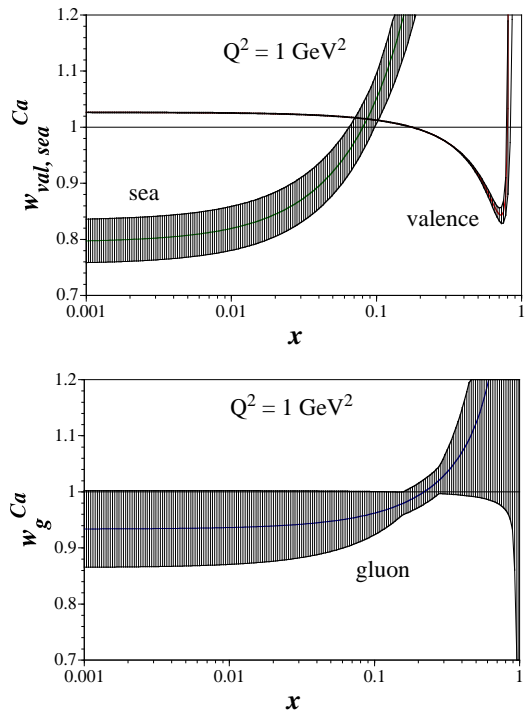}
\vspace{-1.2cm}
\caption{Weight functions with errors for the calcium
         in the quadratic analysis. For an isoscalar nucleus like
         the calcium, the valence up- and down-quark functions
         are the same.}
\label{fig:q-werr}
\end{figure}

\begin{figure}[t!]
\includegraphics[width=0.40\textwidth]{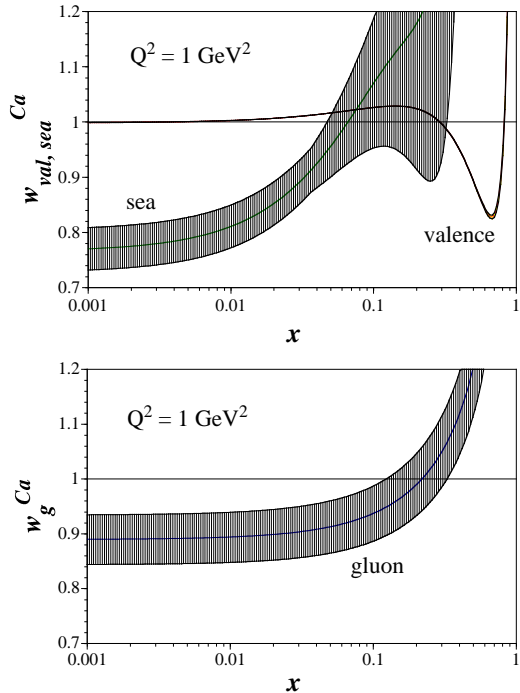}
\vspace{-1.2cm}
\caption{Weight functions with errors for the calcium
         in the cubic analysis.}
\label{fig:c-werr}
\end{figure}

Next, errors are discussed on the obtained weight functions.
As an output of the $\chi^2$ fit by the {\sc Minuit} subroutine,
the optimum parameters and errors are obtained.
The error matrix has a complicated form with nondiagonal
elements. It is not straightforward to perform a rigorous error analysis
with the complicated error matrix. The project is, for example,
in progress as an activity of the Asymmetry Analysis Collaboration
(AAC) \cite{aac}, and there are also recent studies in Ref. \cite{error}.
Here, we employ a simple method which is used 
for example in Ref. \cite{smc}. Effects of the {\sc Minuit}
errors on $w_i$ are calculated exclusively for each parameter,
and then maximum variations are shown as errors in the function $w_i$.
The calcium results are shown in Figs. \ref{fig:q-werr}
and \ref{fig:c-werr} for the quadratic and cubic analyses,
respectively. In an isoscalar nucleus like the calcium,
the valence up-quark function is the same as the valence
down-quark function.
The valence-quark functions have some errors around
the minimum point at $x \sim 0.7$; however,
the small $x$ region is well determined as long as the functional
form is fixed. However, there are uncertainties in the small $x$ behavior
since both valence functions are different at small $x$
in Figs. \ref{fig:q-werr} and \ref{fig:c-werr}. This
kind of error, originating from the assumed functional form, 
is not taken into account in the error bands of these figures.

The antiquark functions have some errors at small $x$;
however, they are obviously shadowed at small $x$. 
The errors and the distribution shapes are very similar at small $x$
in Figs. \ref{fig:q-werr} and \ref{fig:c-werr}. 
Other interesting point is that both errors are very different
in the medium $x$ region, $x>0.1$. Therefore, the antiquark
weight function cannot be well determined at $x>0.1$,
and it depends on the assumed functional form.
There should be also differences in the large $x$ region. 
However, because the antiquark distributions are very small
at $x \gtrsim 0.4$ and they do not contribute to $F_2$
significantly, the large-$x$ antiquark distributions
are not important unless we consider a reaction which is 
sensitive to them.

As shown in Figs. \ref{fig:q-werr} and \ref{fig:c-werr},
the gluon weight functions have large errors in the whole $x$ region.
The first reason for the large errors
is that the analyses are done in the leading order,
and the second is that only $F_2$ data are used for the
$\chi^2$ analyses. Nevertheless, it is interesting to find
that the gluon distributions are shadowed at small $x$
even if the errors are taken into account. Next, there is a
tendency of increase as $x$ becomes larger. Determination
of large-$x$ gluon distributions is not possible in the present
analyses. 
From the simple estimate, we showed the errors in the weight 
functions. However, these studies are intended to give
rough ideas on the errors.
In future, we try to investigate a more complete error analysis.

\begin{figure}[b!]
\vspace{0.3cm}
\includegraphics[width=0.40\textwidth]{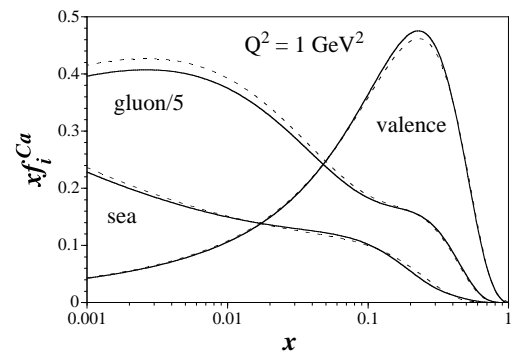}
\vspace{-0.5cm}
\caption{Parton distributions for the calcium are shown.
         The dashed and solid curves indicate
         the quadratic and cubic results, respectively.
         For the calcium, the valence up- and down-quark distributions
         are the same.}
\label{fig:fax}
\end{figure}

Using the results for the weight functions, we show the
parton distributions for the calcium nucleus 
at $Q^2$=1 GeV$^2$ in Fig. \ref{fig:fax}. 
The dashed and solid curves are the quadratic and cubic analysis
results, respectively.
From the $F_2^A$ measurements, the quark distributions are relatively well
determined. For determining the gluon distributions and the details
of the quark distributions, we need to use other reaction data.
Especially, future hadron-collider data should be useful. 

From these analyses, we clarified how well the nuclear parton
distributions can be determined only by the measurements
of the structure functions $F_2^A$.
As mentioned in the introduction, the nuclear parametrization is
still premature in the sense that theoretical and experimental efforts are
necessary for determining the accurate distributions. As far as
the parametrization fit is concerned, much detailed analyses should
be done as an extension of our present studies. On the other hand,
the authors hope that experimental efforts will be made for probing
the valence-quark distributions at small $x$ by a neutrino factory
and for finding the antiquark and gluon distributions by 
hadron colliders such as RHIC.

Our studies should be also important for investigating the parton
distributions in the nucleon. As mentioned in Sec. \ref{intro},
nuclear data have been used partially for obtaining the parton
distributions in the nucleon. In particular, neutrino data are
important for determining the valence-quark distributions;
however, the data are taken for example for the iron nucleus.
We need to feed back our studies to readjust the distributions
in the ``nucleon".

\section{Practical nuclear parton distributions}
\label{usage}

The nuclear parton distributions obtained in our analyses could be used
for studying other high-energy nuclear reactions.
We propose two types of distributions which are obtained in the
quadratic and cubic type analyses. Because the $\chi_{min}^2$ is smaller
for the cubic type, we prefer it to the quadratic type.
The distributions are provided either in the analytical form
at $Q^2$=1 GeV$^2$ or in the form of computer subroutines.
Although there could exist distributions at $x>1$ in a nucleus,
such large $x$ distributions are not provided in our studies
as explained in Sec. \ref{x-depend}.
The analytical expressions are given in Sec. \ref{analytical},
the subroutines are explained in Sec. \ref{library}.

\subsection{Analytical expressions}
\label{analytical}

The analytical expressions are given at $Q^2$=1 GeV$^2$. 
Therefore, one needs to evolve the distributions to a specific
$Q^2$ point by one's own evolution code with
$\Lambda_{LO}^{MRST}=0.1741$ GeV.
If it were the case where $Q^2$ dependence can be neglected,
one may use the analytical distributions without the evolution. 
The nuclear distributions should be calculated by Eq. (\ref{eqn:wpart})
with the obtained weight functions and MRST-LO (central gluon)
distributions \cite{mrst}. However, one should be careful
that the antiquark distributions are
slightly modified from the original form 
so as to become flavor symmetric $\bar q \equiv Sea(MRST)/6$,
because the antiquark flavor asymmetry is not taken into
account in our nuclear analyses.

\subsubsection{Type \uppercase\expandafter{\romannumeral 1}:
               Cubic fit}

We call the cubic distributions
type \uppercase\expandafter{\romannumeral 1} distributions.
The weight functions obtained in the cubic fit are
\begin{align}
w_{u_v}  & = 1+\left( 1 - \frac{1}{A^{1/3}} \right) 
\nonumber \\ & \ \ \ \ \ \ \ \ \times
          \frac{a_{u_v}(A,Z) +0.6222 x - 2.858 x^2 
                             +2.557 x^3}{(1-x)^{0.8107}} ,
\nonumber \\
w_{d_v}  & = 1+\left( 1 - \frac{1}{A^{1/3}} \right) 
\nonumber \\ & \ \ \ \ \ \ \ \ \times
          \frac{a_{d_v}(A,Z) +0.6222 x - 2.858 x^2
                             +2.557 x^3}{(1-x)^{0.8107}} ,
\nonumber \\
w_{\bar q} & = 1+\left( 1 - \frac{1}{A^{1/3}} \right) 
\nonumber \\ & \ \ \ \ \ \ \ \ \times
          \frac{-0.3313 +6.995 x -34.17 x^2 
                                 +62.54 x^3}{1-x} ,
\nonumber \\
w_{g}      & = 1+\left( 1 - \frac{1}{A^{1/3}} \right) 
          \frac{a_{g}(A,Z) +0.8008 x -0.4004 x^2}{1-x} .
\end{align}
The nuclear dependent constants are listed in Table \ref{tab:c-audg}
for the used nuclei. They depend on the mass number $A$ and atomic
number $Z$ in general.

\begin{table}[h]
\caption{Obtained parameters $a_{u_v}$, $a_{d_v}$, and $a_{g}$ 
         for the used nuclei in the cubic analysis.}
\label{tab:c-audg}
\begin{ruledtabular}
\begin{tabular*}{\hsize}
{c@{\extracolsep{0ptplus1fil}}c@{\extracolsep{0ptplus1fil}}c
@{\extracolsep{0ptplus1fil}}c}
nucleus & $a_{u_v}$ & $a_{d_v}$ & $a_{g}$ \\
\colrule
D     &  $-$0.002178  &  $-$0.002178  &  $-$0.1560 \    \\
He    &  $-$0.002178  &  $-$0.002178  &  $-$0.1560 \    \\
Li    &  $-$0.002690  &  $-$0.001716  &  $-$0.1560 \    \\
Be    &  $-$0.002571  &  $-$0.001815  &  $-$0.1560 \    \\
C     &  $-$0.002178  &  $-$0.002178  &  $-$0.1560 \    \\
N     &  $-$0.002178  &  $-$0.002178  &  $-$0.1560 \    \\
Al    &  $-$0.002306  &  $-$0.002054  &  $-$0.1560 \    \\
Ca    &  $-$0.002178  &  $-$0.002178  &  $-$0.1560 \    \\
Fe    &  $-$0.002427  &  $-$0.001941  &  $-$0.1560 \    \\
Cu    &  $-$0.002456  &  $-$0.001916  &  $-$0.1560 \    \\
Ag    &  $-$0.002610  &  $-$0.001782  &  $-$0.1560 \    \\
Sn    &  $-$0.002726  &  $-$0.001686  &  $-$0.1560 \    \\
Xe    &  $-$0.002814  &  $-$0.001616  &  $-$0.1560 \    \\
Au    &  $-$0.002902  &  $-$0.001549  &  $-$0.1560 \    \\
Pb    &  $-$0.002955  &  $-$0.001509  &  $-$0.1560 \    \\	
\end{tabular*}
\end{ruledtabular}
\end{table}

As obvious from the table, there are significant nuclear dependence
in the parameters $a_{u_v}$ and $a_{d_v}$. However,
the dependence is so small in the parameter $a_{g}$
that it cannot be shown in the table.
These parameters are the same for isoscalar nuclei
because of the conditions in Eqs. (\ref{eqn:charge}), (\ref{eqn:baryon}),
and (\ref{eqn:momentum}), and this fact is clearly shown in
Appendix \ref{appen-a}.
If one would like to have analytical expressions for a nucleus
which is not listed in Table \ref{tab:c-audg}, there are following
two possibilities.
The first method is that one calculates $a_{u_v}$, $a_{d_v}$,
and $a_{g}$ so as to satisfy the conditions of nuclear charge,
baryon number, and momentum in Eqs. (\ref{eqn:charge}), (\ref{eqn:baryon}),
and (\ref{eqn:momentum}) for one's chosen nucleus. 

For those who think this calculation is tedious, we prepare
an alternative method. Using the three conditions,
we find that $a_{u_v}$, $a_{d_v}$, and $a_{g}$ can be expressed
in terms of eight integrals, which are nuclear independent, together
with $A$ and $Z$. The details of this method are so technical that
they are discussed in Appendix \ref{appen-a}.
If one still thinks that these calculations are too much works to do,
or if one does not have a $Q^2$ evolution subroutine,
one had better use computer codes explained in Sec. \ref{library}
for getting numerical values of the parton distributions.

\subsubsection{Type \uppercase\expandafter{\romannumeral 2}:
               Quadratic fit}

We call the quadratic distributions
type \uppercase\expandafter{\romannumeral 2} distributions.
The weight functions obtained in the quadratic fit are
\begin{align}
w_{u_v}    & = 1+\left( 1 - \frac{1}{A^{1/3}} \right) 
          \frac{a_{u_v}(A,Z) -0.2593 \, x + 0.2586 \, x^2}{(1-x)^{2.108}} ,
\nonumber \\
w_{d_v}    & = 1+\left( 1 - \frac{1}{A^{1/3}} \right) 
          \frac{a_{d_v}(A,Z) -0.2593 \, x + 0.2586 \, x^2}{(1-x)^{2.108}} ,
\nonumber \\
w_{\bar q} & = 1+\left( 1 - \frac{1}{A^{1/3}} \right) 
          \frac{-0.2900 +3.774 \, x -2.236 \, x^2}{1-x} ,
\nonumber \\
w_{g}      & = 1+\left( 1 - \frac{1}{A^{1/3}} \right) 
          \frac{a_{g}(A,Z) + 0.4798 \, x -0.2399 \, x^2}{1-x} .
\end{align}
The nuclear dependent parameters are listed in Table \ref{tab:q-audg}.
We find that the $A$ dependent variations are very small
in these parameters of the quadratic fit.
If one needs expressions for other nucleus, one should evaluate
$a_{u_v}$, $a_{d_v}$, and $a_{g}$ as suggested in the type I section.

\begin{table}[h]
\caption{Obtained parameters $a_{u_v}$, $a_{d_v}$, and $a_{g}$
          for the used nuclei in the quadratic analysis.}
\label{tab:q-audg}
\begin{ruledtabular}
\begin{tabular*}{\hsize}
{c@{\extracolsep{0ptplus1fil}}c@{\extracolsep{0ptplus1fil}}c
@{\extracolsep{0ptplus1fil}}c}
nucleus & $a_{u_v}$ & $a_{d_v}$ & $a_{g}$ \\
\colrule
D     &  0.03745  &  0.03745  &  $-$0.09391 \    \\
He    &  0.03745  &  0.03745  &  $-$0.09391 \    \\
Li    &  0.03709  &  0.03776  &  $-$0.09392 \    \\
Be    &  0.03717  &  0.03770  &  $-$0.09392 \    \\
C     &  0.03745  &  0.03745  &  $-$0.09391 \    \\
N     &  0.03745  &  0.03745  &  $-$0.09391 \    \\
Al    &  0.03736  &  0.03753  &  $-$0.09391 \    \\
Ca    &  0.03745  &  0.03745  &  $-$0.09391 \    \\
Fe    &  0.03727  &  0.03761  &  $-$0.09391 \    \\
Cu    &  0.03725  &  0.03763  &  $-$0.09391 \    \\
Ag    &  0.03714  &  0.03772  &  $-$0.09392 \    \\
Sn    &  0.03706  &  0.03778  &  $-$0.09392 \    \\
Xe    &  0.03700  &  0.03783  &  $-$0.09393 \    \\
Au    &  0.03694  &  0.03788  &  $-$0.09394 \    \\
Pb    &  0.03690  &  0.03790  &  $-$0.09394 \    \\
\end{tabular*}
\end{ruledtabular}
\end{table}

\subsection{Parton distribution library}
\label{library}

If one needs to have nuclear parton distributions in a numerical form
at a given $x$ and $Q^2$ point, one may use the computer codes
in Ref. \cite{nucl-lib}.
Two kinds of subroutines are made. First, there is a subroutine
for the used nuclei in this paper. For other nuclei, we prepared 
the second one.

First, if one wishes to calculate the distributions in the used nuclei:
D, $^4$He, Li, Be, C, N, Al, Ca, Fe, Cu, Ag, Sn, Xe, Au, and Pb,
one should use the first code. In addition, we prepared 
the distributions in the nucleon because we modified the MRST
antiquark distributions as flavor symmetric in our studies.
The kinematical ranges are $10^{-9} \le x \le 1$ and
$1 \ {\rm GeV}^2 \le Q^2 \le 10^5 \ {\rm GeV}^2$.
The variables $x$ and $Q^2$ are divided into small steps.
Then, a grid data set is prepared for the parton distributions
in each nucleus. Because the scaling violation is a rather
small effect, a simple linear interpolation in $log \, Q^2$ is used for
calculating the distributions at a given $Q^2$.
On the other hand, because the $x$ dependence is more complicated,
a cubic Spline interpolation is used for calculating 
the distributions at a given $x$ point. Running this code,
one obtains the distributions, $x u_v^A$, $x d_v^A$, $x \bar q ^A$,
and $x g^A$, for a specified nucleus at a given $x$ and $Q^2$ point.
Even though the antiquark distributions are flavor symmetric 
at $Q^2$=1 GeV$^2$, they are not symmetric at different $Q^2$
in the next-to-leading order \cite{skpr}. However, because 
such $Q^2$ evolution effects do not exist in the leading order,
the antiquark distributions are consistently flavor symmetric
at any $Q^2$.

Second, if one would like to have the distributions in other nucleus,
one should use the second code. Here, the analytical expressions
in Sec. \ref{analytical} are used as the initial distributions.
At first, the constants $a_{u_v}$, $a_{d_v}$, and $a_{g}$ are calculated
so as to satisfy the charge, baryon-number, and momentum conditions
for a given nucleus with $A$ and $Z$.
Then, they are evolved to a requested $x$ and $Q^2$ point
by the ordinary DGLAP evolution equations in Ref. \cite{bf1}. 
However, one has to be careful about the requested nucleus in the sense
that it should not be too far away from the used nuclei. For example,
we do not support the distributions in an extremely unstable nucleus
with large neutron excess. Strictly speaking, huge nuclei with $A>208$
are also outside our supporting range. However, as obvious from
Figs. \ref{fig:q-wxa} and \ref{fig:c-wxa}, the variations of the 
parton distributions are already very small between the calcium
with $A=40$ and gold with $A=197$, so that the extrapolation
from $A=208$ to nuclear matter is {\it expected} to be reliable.
The details of the usage are explained in Ref. \cite{nucl-lib}.

In the second code, it takes time for getting the results because
the $Q^2$ evolution calculations consume computing time.
It does not matter to calculate the distributions for a few
$Q^2$ points. However, if one would like to use it frequently,
one may try the following. The second code is prepared so that
one could create a grid file for a requested nucleus. Then, one
can use it in the first code, where the computation is much faster.

\section{Summary}\label{sum}

We have done the global analyses of existing experimental data on nuclear
$F_2$ for obtaining optimum parton distributions in nuclei.
Assuming a simple yet reasonable overall $A$ dependence, the nuclear
parton distributions are expressed in terms of a number of parameters.
The quadratic and cubic functional forms are assumed for the $x$ dependence.
The parameters have been determined by the $\chi^2$ analyses.
As a result, we obtained reasonable fit to the measured experimental data
of $F_2$. The valence-quark distributions are relatively well determined
except for the fact that the small $x$ part depends slightly on the assumed
functional form. The antiquark distributions are reasonably well determined
at small $x$; however, the large $x$ behavior is not obvious from
the $F_2$ data. The analyses indicated that the gluon distributions
are shadowed at small $x$: however, they cannot be well determined
by the present $F_2$ data, especially in the large $x$ region. 

We have proposed two types of nuclear parton distributions which are
obtained by the quadratic and cubic type analyses. They are provided
either by the analytical expressions at $Q^2$=1 GeV$^2$ or by 
the computer programs for calculating them numerically.
Our analyses should be important not only for understanding physics
mechanisms of nuclear modification but also for applications to
heavy-ion physics. Our results could also shed light on an issue of
present parton distributions in the nucleon because nuclear data
have been partially used in the ``nucleon" analysis.

\begin{acknowledgments}
All the authors were supported by the Grant-in-Aid for Scientific Research
from the Japanese Ministry of Education, Culture, Sports, Science,
and Technology. M.H. and M.M. were supported by the JSPS Research Fellowships
for Young Scientists. They thank the AAC members for discussions
on general techniques of $\chi^2$ analysis. 
\end{acknowledgments}

\appendix
\section{Nuclear dependent parameters}
\label{appen-a}

From the conditions of nuclear charge, baryon number, and momentum,
the nuclear dependent parameters, $a_{u_v}$, $a_{d_v}$, and $a_{g}$,
can be expressed in terms of eight integrals together with $A$ and $Z$.
It is the advantage that these integrals are nuclear independent.
Therefore, reading the numerical values of the integrals
and using the equations in this section, one can easily
calculate the values of $a_{u_v}$, $a_{d_v}$, and $a_{g}$ for any
nuclei. The necessary integrals are the following:
\begin{align}
I_1    & = \int dx \frac{H_v(x)}{(1-x)^{\beta_v}} \, u_v(x) , \ \ \ \ 
I_2      = \int dx \frac{H_v(x)}{(1-x)^{\beta_v}} \, d_v(x) ,
\nonumber \\
I_3    & = \int dx \frac{1}{(1-x)^{\beta_v}} \, u_v(x) ,      \ \ \ \ 
I_4      = \int dx \frac{1}{(1-x)^{\beta_v}} \, d_v(x) ,
\nonumber \\
I_5    & = \int dx \frac{x}{(1-x)^{\beta_v}} \, u_v(x) ,      \ \ \ \ 
I_6      = \int dx \frac{x}{(1-x)^{\beta_v}} \, d_v(x) ,
\nonumber \\
I_7    & = \int dx \, x \, \bigg[ \, \frac{H_v(x)}{(1-x)^{\beta_v}} \, 
                                     \{ u_v(x)+d_v(x) \}
\nonumber \\
    & \ \ \ \ \ \ \ \ \ \ \ \ \ \ \ 
                + \frac{a_{\bar q} +H_{\bar q}(x)}{1-x} \, 6 \, \bar q(x)
                + \frac{H_g(x)}{1-x} \, g(x)  \, \bigg] ,
\nonumber \\
I_8    & = \int \frac{x}{1-x} \, g(x) .
\label{eqn:8int}
\end{align}
We should note that the parton distributions in these equations are
those in the nucleon. The functions $H_i(x)$ are the one given
in Eq. (\ref{eqn:wi-withh}). 
The integral values are numerically given in Table \ref{tab:a-values}.

\begin{table}[h!]
\caption{Values of the integrals are given.}
\label{tab:a-values}
\begin{ruledtabular}
\begin{tabular*}{\hsize}
{c@{\extracolsep{0ptplus1fil}}c@{\extracolsep{0ptplus1fil}}c}
Integral & Type \uppercase\expandafter{\romannumeral 1}
         & Type \uppercase\expandafter{\romannumeral 2} \\
\colrule
$I_1$    &  $-$0.0007990     &   $-$0.1474    \\
$I_2$    &     0.008540      &   $-$0.05297   \\
$I_3$    &     2.406         &      3.768     \\
$I_4$    &     1.148         &      1.583     \\
$I_5$    &     0.4777        &      1.157     \\
$I_6$    &     0.1772        &      0.3629    \\
$I_7$    &     0.08326       &   $-$0.007650  \\
$I_8$    &     0.5246        &      0.5246    \\
\end{tabular*}
\end{ruledtabular}
\end{table}

Using these integrals, we can express
the nuclear dependent parameters as
\begin{align}
a_{u_v}(A,Z) & = - \frac{Z I_1 + (A-Z) I_2}{Z I_3 + (A-Z) I_4} ,
\nonumber \\
a_{d_v}(A,Z) & = - \frac{Z I_2 + (A-Z) I_1}{Z I_4 + (A-Z) I_3} ,
\nonumber \\
a_{g}(A,Z)   & = - \frac{1}{I_8} \,  \bigg[ \,
            a_{u_v}(A,Z) \left\{ \frac{Z}{A} I_5
                                +\left(1-\frac{Z}{A} \right) I_6 \right\}
\nonumber \\
         & \! \! \! \! \! \! \! 
           +a_{d_v}(A,Z) \left\{ \frac{Z}{A} I_6
                                +\left(1-\frac{Z}{A} \right) I_5 \right\}
           +I_7 \, \bigg]   .
\label{eqn:a-int}
\end{align}
From Table \ref{tab:a-values} and Eq. (\ref{eqn:a-int}),
it is possible to calculate the parton distributions in any nucleus.
However, we recommend to use our results for a nucleus
which is rather close to the analyzed nuclei.



\end{document}